\documentclass[aps,prd,preprintnumbers,superscriptaddress,nofootinbib]{revtex4-2} 
\usepackage[utf8]{inputenc}
\usepackage{amssymb,amsmath,amsfonts}
\usepackage{multirow}
\usepackage{slashed} 
\usepackage{graphicx}
\usepackage{subfigure}
\usepackage{hyperref}
\hypersetup{colorlinks=true,linkcolor=purple,anchorcolor=blue,citecolor=blue, filecolor=blue,urlcolor=red,bookmarksnumbered=true,
	pdfview=FitB
}

\usepackage{color}
\usepackage{xcolor}
\usepackage{ulem}
\usepackage{csquotes}
\colorlet{purple1}{blue!70!red}
\colorlet{darkred}{red!50!black}
\usepackage{float}

\def\orcid#1{\kern .08em\href{https://orcid.org/#1}{\includegraphics[keepaspectratio,width=0.7em]{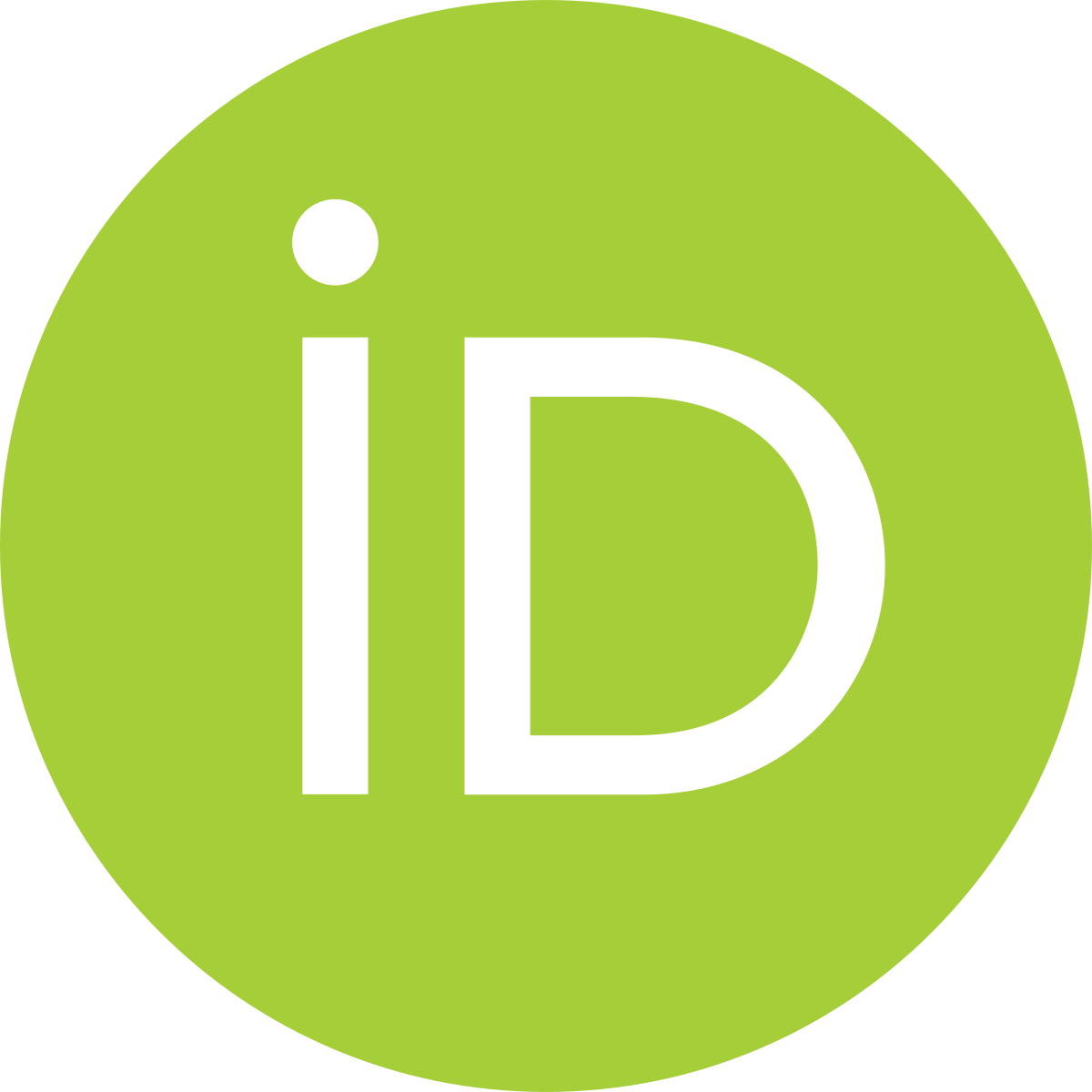}}}

\def\orcid#1{\kern .08em\href{https://orcid.org/#1}{\includegraphics[keepaspectratio,width=0.7em]{ORCID_iD.png}}}

\begin{document}

	%	\title{Diffractive $\rho$-meson production at HERA using a holographic AdS/QCD light-front meson wave function and the ’t Hooft equation}

  %	\title{Diffractive $\rho$-meson production at HERA using the holographic light-front Schrödinger equation and the ’t Hooft equation}

  	\title{$\rho$-meson spectroscopy and diffractive production using the holographic light-front
Schrödinger equation and the ’t Hooft equation}

		\author{Bheemsehan~Gurjar\orcid{0000-0001-7388-3455}}
		\email{gbheem@iitk.ac.in} 
		\affiliation{Department of Physics, Indian Institute of Technology Kanpur, Kanpur-208016, India}
%		\item \href{https://orcid.org/0000-0000-0000-0000}{\textcolor{orcidlogocol}{\aiOrcid} \hspace{2mm} orcid.org/0000-0000-0000-0000}
     	\author{Chandan~Mondal\orcid{0000-0002-0000-5317}}
		\email{mondal@impcas.ac.cn}
		\affiliation{Institute of Modern Physics, Chinese Academy of Sciences, Lanzhou 730000, China}
		\affiliation{School of Nuclear Science and Technology, University of Chinese Academy of Sciences, Beijing 100049, China}
		
		\author{Satvir~Kaur\orcid{0000-0002-7643-5970}}
		\email{satvir@impcas.ac.cn}
		\affiliation{Institute of Modern Physics, Chinese Academy of Sciences, Lanzhou 730000, China}
		\affiliation{School of Nuclear Science and Technology, University of Chinese Academy of Sciences, Beijing 100049, China}
\begin{abstract}

We determine the mass spectroscopy and light-front wave functions (LFWFs) of the $\rho$-meson by solving the holographic Schrödinger equation of light-front chiral QCD along with the ’t Hooft equation of (1+1)-dimensional QCD in the large $N_c$ limit. Subsequently, we utilize the obtained LFWFs in conjunction with the color glass condensate dipole cross-section to calculate the cross sections for the diffractive $\rho$-meson electroproduction. Our spectroscopic results align well with the experimental data. Predictions for the diffractive cross sections demonstrate good consistency with the available experimental data at different energies from H1 and ZEUS collaborations. Additionally, we show that the resulting LFWFs for the $\rho$-meson can effectively describe various properties, including its decay constant,   distribution amplitudes, electromagnetic form factors, charge radius, magnetic and quadrupole moments. Comparative analyses are conducted with experimental measurements and the available theoretical predictions.

\end{abstract}	
\maketitle
%
%%%%%%%%%%%%%%%%%
\section{Introduction}
%%%%%%%%%%%%%%%%%
Experimental processes such as deep inelastic scattering (DIS), deeply virtual Compton scattering (DVCS), and exclusive diffractive vector meson production serve as effective tools to investigate Quantum Chromodynamics (QCD)~\cite{Gribov:1983ivg}. Particularly, at small $x$, these processes are predominantly influenced by gluon saturation. This phenomenon has extensively been explored within the framework of the Color Glass Condensate (CGC) effective field theory~\cite{Nikolaev1991, MUELLER1994471, JALILIANMARIAN2006104, McLerran:2001sr, Iancu:2002xk}. The CGC theory describes the balance of gluons through recombination and multiple scattering limitations within a dipole picture~\cite{Mueller:1993rr,Nemchik:1996cw, NEMCHIK1996199,NEMCHIK1994228}. In this dipole model, a virtual photon splits into a quark and an anti-quark pair (dipole), interacts with proton through gluon exchange, and reforms into a vector meson or photon as illustrated in Fig.~\ref{fig:CGCmodel}.

The goal of this paper is to  predict the cross section for diffractive $\rho$-meson %and $\phi$ 
electroproduction, observed at the HERA collider~\cite{H1:1999pji, H1:2009cml, H1:1996prv, ZEUS:2005bhf,ZEUS:2007iet,ZEUS:1997rof}, using the QCD color dipole model and the nonperturbative holographic meson light-front wave functions (LFWFs)~\cite{BRODSKY20151} by taking into account the longitudinal dynamics generated by the 't Hooft Equation in $(1 + 1)$-dim QCD at large $N_c$~\cite{tHooft:1974pnl}.

The holographic light-front QCD (hLFQCD) is developed within the chiral limit of light-front QCD, establishing an exact correspondence between strongly coupled (1+3)-dimensional light-front QCD and weakly interacting string modes in (1+4)-dimensional anti-de-Sitter (AdS) space. For a review of hLFQCD, see Ref. \cite{Brodsky:2014yha}.
The primary nontrivial prediction of this approach leads to the lightest bound
state, i.e., the pion is massless. Another crucial prediction asserts that the meson masses align along the universal Regge trajectories, mirroring experimental observations. Note that the predicted slopes are dictated by the strength of the confining potential, $\kappa$. The form of the confining potential in physical spacetime is determined by a dilaton field that breaks the conformal symmetry of AdS space. A phenomenologically successful choice involves a quadratic dilaton in the fifth dimension of AdS space, which corresponds to a light-front harmonic oscillator in physical spacetime. The mass scale parameter, $\kappa$, is fixed by fitting the experimentally observed slopes of the meson mass spectrum Regge trajectories for different meson groups. It is found that for all the light mesons, $\kappa \simeq 0.5$ GeV~\cite{BRODSKY20151}.

Going beyond the semiclassical approximation, Brodsky and de Téramond proposed an invariant mass ansatz (IMA) for including nonzero quark masses~\cite{Brodsky:2008pg}. Using IMA, one can calculate the shift in the meson masses as a first order perturbation. The predicted mass shift for the pion (and kaon) appears as same as the physical mass of the meson. These results have been obtained by fixing the scale parameter as $\kappa = 0.54$ GeV, and the light quark masses as $m_{u/d} = 0.046$ GeV and $m_{s} = 0.357$ GeV, which vanish in the chiral limit~\cite{BRODSKY20151}.

Previous works~\cite{PhysRevLett.109.081601,PhysRevD.94.074018} reported predictions for vector mesons production by utilizing the holographic wave function with IMA together with the CGC dipole cross section~\cite{PhysRevD.78.014016}.
Reference~\cite{PhysRevLett.109.081601} has investigated the process of $\rho$-meson production using a light quark mass of $m_{q} = 0.14$ GeV, which is consistent with the fitted parameters of CGC dipole cross section~\cite{PhysRevD.74.074016,Jeffrey} from the inclusive DIS data~\cite{ZEUS:2001mhd,H1:2000muc}. 
The most recent analyses of dipole cross sections have utilized the 2010 DIS data at HERA~\cite{H1:2009pze}. It has been acknowledged in Ref.~\cite{PhysRevD.78.014016} that the DIS data prefers the lower light quark masses, but also noted that the use of effective quark mass, $m_{q} = 0.14$ GeV, yields satisfactory fit to the 2001 DIS structure function data. In a more recent study~\cite{Contreras:2015joa}, a novel dipole model has demonstrated that both current quark mass and effective quark mass $m_{q} = 0.14$ GeV accurately fit the 2010 DIS structure function data~\cite{H1:2009pze}. In Ref.~\cite{PhysRevD.94.074018}, %\sout{the authors updated CGC dipole model parameters with light quark masses derived from the conclusive 2015 HERA data on inclusive DIS.} 
{the authors revisited the CGC dipole model and fit the conclusive 2015 HERA data of inclusive DIS with the light quark masses.} They studied cross sections for the diffractive $\rho$ and $\phi$ meson production using the fitted dipole cross section~\cite{IANCU2004199}, the perturbatively calculated photon LFWFs~\cite{PhysRevD.22.2157,PhysRevD.55.2602}, and the holographic meson LFWFs with IMA, which contains no dynamical information of the meson in the longitudinal direction~\cite{PhysRevLett.102.081601}. 

In this work, the nonzero light quark masses are incorporated through the chiral symmetry breaking and the longitudinal dynamics governed by the ’t Hooft equation of (1+1)-dimensional QCD in the large $N_c$ limit. The combined holographic Schrödinger equation and the 't Hooft equation provide a reliable picture of the data to the $\rho$-meson spectrum with the universal $\kappa$. %~\cite{Ahmady:2021yzh}. Here,
We show that, together, they can simultaneously describe various properties of the $\rho$-meson,
including its decay constant, parton distribution amplitude (PDA), electromagnetic form factors, charge radius etc., as well as, in conjunction
with the color glass condensate dipole cross-section, the resulting wave functions can provide good description of the HERA data of the diffractive $\rho$-meson electroproduction.

The rest of the paper is organized as follows: In Sec.~\ref{CGCmodel}, we review the color dipole model. The holographic meson LFWFs %light-front wave functions  
followed by the longitudinal dynamics using the 't Hooft equation are discussed in Sec.~\ref{sec:HLFQCD}. In Sec.~\ref{sec:results}, we discuss the numerical results for the mass spectroscopy, the diffractive cross sections using the dipole cross section, PDAs,  electromagnetic form factors, decay constant, charge radius, magnetic and quadrupole moments for the $\rho$-meson with the holographic meson wave function. Finally, we conclude the paper in Sec.~\ref{sec:conclusion}.
%
%%%%%%%%%%%%%%%%%
\section{THE DIPOLE MODEL of EXCLUSIVE VECTOR MESON PRODUCTION}\label{CGCmodel}
\begin{figure}
	\centering
 	\includegraphics[width=\textwidth]{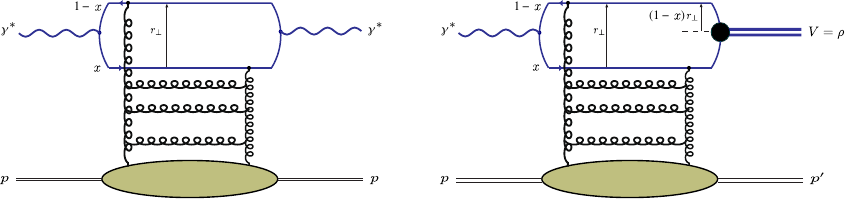}
	\caption{ The $\gamma^{\ast}p$ elastic scattering amplitude for DIS (left) and for exclusive vector meson production (right) in dipole model. Here, $x$ is the fraction of the photon momentum carried by the quark and $\mathbf{r}_{\perp}$ is the transverse size of the dipole.}% and $\vec{b}$ is the impact parameter of the dipole.}
	\label{fig:CGCmodel}
\end{figure}

%%%%%%%%%%%%%%%%%
In the dipole picture, the scattering amplitude for the diffractive process $\gamma^\ast p \rightarrow V p$ is expressed as the convolution of the overlap of the LFWFs of the photon  and the vector meson, and the proton-dipole scattering amplitude~\cite{PhysRevD.74.074016},
%Mathematically, the scattering amplitude for the exclusive vector meson production in DIS is formulated as~\cite{PhysRevD.74.074016},
\begin{eqnarray}
\Im \mbox{m}\, \mathcal{A}^{\gamma^{\ast}p\rightarrow Vp}_\Lambda(s,t;Q^2)  
 &=&   \sum_{h, \bar{h}} \int {\mathrm d}^2 {\mathbf r}_{\perp} \; {\mathrm d} x \; \Psi^{\gamma^*,\Lambda}_{h, \bar{h}}(x,{\mathbf r}_{\perp};Q^2)  \Psi^{V,\Lambda}_{h, \bar{h}}(x,{\mathbf r}_{\perp})^* e^{-i x \mathbf{r}_{\perp} \cdot \mathbf{\Delta}} \mathcal{N}(x_{\text{m}},\mathbf{r}_{\perp}, \mathbf{\Delta}),
\label{eq:amplitude-VMP} 
\end{eqnarray}
where $Q^{2}$ is the virtuality of the photon and  $t=-\Delta^{2}$ is the squared of the transverse momentum transfer at the proton vertex. %{\color{red} in transverse direction?}. 
The substantial value of the center-of-mass energy squared, $s$, ensures the factorization of the scattering amplitude for the diffractive process  into a convolution of the LFWFs of the photon, $\Psi^{\gamma,\Lambda}_{h, \bar{h}}(x,{\mathbf r}_{\perp};Q^2)$, and vector meson, $\Psi^{V,\Lambda}_{h, \bar{h}}(x,{\mathbf r}_{\perp})$, and a dipole cross section, $\mathcal{N}(x_m,{\mathbf r}_{\perp},\Delta)$. 
The variable ${\mathbf r}_{\perp}$ is the transverse size of the dipole and $x$ defines the longitudinal momentum fraction carried by the quark as shown in Fig.~\ref{fig:CGCmodel}.  
 The indices $h$ and $\bar{h}$ are the helicities of the quark and the antiquark.
 The symbol $\Lambda$ in the superscript of the LFWFs denotes the polarization of the photon and the vector meson. The systems can be longitudinally polarized or transversely polarized; symbolically, $\Lambda=0~\textrm{or}~\pm 1$, respectively. 
  The dipole-proton scattering amplitude depends upon the center-of-mass energy of the photon-proton system ($W$), %{\sout{defined through}}
  {related to} the modified Bjorken variable {as}~\cite{Rezaeian:2013tka}:  ${x_{\rm m}=x_{\rm Bj}\left(1+\frac{M_{V}^{2}}{Q^{2}}\right)}$, where $x_{\rm Bj}=\frac{Q^{2}}{W^{2}}$. The dipole-proton scattering amplitude encapsulates the high-energy QCD dynamics associated with the dipole-proton interaction. Being a universal object, it can be obtained via an approximate solution
of the Balitsky-Kovchegov equation~\cite{Balitsky:1995ub,PhysRevD.60.034008,PhysRevD.61.074018} within the CGC formalism~ \cite{Jalilian-Marian:1997qno,PhysRevD.59.014014,Iancu:2000hn,Iancu:2001ad,Weigert:2000gi}.

The differential cross section for the exclusive vector meson production is given by~\cite{PhysRevD.78.014016,ZEUS:2007iet},
\begin{eqnarray}\label{eq:diffcrosssect} 
	\frac{d\sigma_{\Lambda}^{\gamma^{\ast}p\rightarrow Vp}}{dt}= \frac{1}{16\pi} 
	[\Im\mathrm{m} \mathcal{A}^{\gamma^{\ast}p\rightarrow Vp}_\Lambda(s, t=0)]^2 \; (1 + \beta_\Lambda^2) \exp(-B_D t)
\end{eqnarray}
with the parameter $\beta_{\Lambda}$ being the ratio of the real to imaginary components of the scattering amplitude expressed as~\cite{PhysRevD.78.014016,PhysRevD.87.034002},
\begin{eqnarray}
\beta_{\Lambda}=\tan\left(\frac{\pi}{2}\alpha_{\Lambda}\right)~~~~~~\textrm{with}~~~~~~\alpha_{\Lambda}=\frac{\partial \textrm{log}|\Im\mathrm{m} \mathcal{A}_{\Lambda}|}{\partial \textrm{log}(1/x_{m})},
\end{eqnarray}
%{\color{red} Is it $x$ or $x_m$?} {\color{blue}It is only $x$ not $x_{m}$}
and the diffractive slope parameter $B_{D}$ parameterized as~\cite{PhysRevD.94.074018},
\begin{eqnarray}
	B_D = N\left[ 14.0 \left(\frac{1~\mathrm{GeV}^2}{Q^2 + M_{V}^2}\right)^{0.2}+1\right]
	\label{eq:Bslope}
\end{eqnarray}
with $N=0.55$ GeV$^{-2}$.
%As mentioned in Ref.~\cite{PhysRevD.94.074018}, 
The parametrization of this slope parameter is consistent with the ZEUS data for $\rho$-meson production~\cite{ZEUS:2007iet,PhysRevD.87.034002}. However, the most recent H1 data~\cite{H1:2009cml} favours somewhat larger value of $B_{D}$ accompanied by the large uncertainty.  %$\beta_\lambda$ in Eq.~(\ref{eq:diffcrosssect}) is a ratio between the real and imaginary parts of the amplitude~\cite{PhysRevD.74.074016,PhysRevD.87.034002}.

A simplified model for the dipole cross-section was proposed a long ago in Ref.~\cite{IANCU2004199}, known as the CGC dipole
model. The dipole cross section is given by%~~~~~~~~~~{\color{red} tilde{N}??}
\begin{eqnarray}
	\hat{\sigma}(x_{\text{m}},r_{\perp}) = \sigma_0 \, { {\mathcal{N}} (x_{\text{m}}, r_{\perp} Q_s, 0 ) }
	\label{eq:sigmaCGC}
\end{eqnarray}
with
\begin{eqnarray}
	{\mathcal{N}} (x_{\text{m}},r_{\perp} Q_s, 0) &=& \begin{cases}
 { \mathcal N}_0 \left ( { r_{\perp} Q_s \over 2 }\right)^ {2 \left [ \gamma_s + { {\mathrm ln}  (2 / r_{\perp} Q_s) \over  \kappa_0 \, \lambda \, {\mathrm ln} (1/x_{\text{m}}) }\right]}  ~~~~~~~{\rm for }~~~~~~~r_{\perp} Q_s \leq 2 \,,\\
{ 1- \exp[-{\mathcal A} \,{\mathrm ln}^2 ( {\mathcal B} \, r_{\perp} Q_s)]}~~~~~~~~~~~~{\rm for }~~~~~~~r_{\perp} Q_s > 2\,
	\end{cases}
	\label{eq:dipolcross}
\end{eqnarray}
where $r_{\perp}=|{\mathbf r}_{\perp}|$ and the saturation scale is $Q_{s}=(x_{0}/x_{\rm m})^{\lambda/2}$ GeV. 
%However Kowalski and Watt proposed a model where the saturation scale have a Gaussian dependence on the impact parameter~\cite{PhysRevD.78.014016}. 
The coefficients $\mathcal{A}$ and $\mathcal{B}$ in Eq.~\eqref{eq:dipolcross} are determined uniquely from the condition that ${\mathcal{N}}(x_{\rm m},r_{\perp}Q_{s},0)$ and its derivative with respect to $r_{\perp}Q_s$ are continuous at $r_{\perp}Q_s=2$. This leads to
\begin{eqnarray} \label{eq:AandB}
 \mathcal{A} = -\frac{\mathcal{N}_0^2\gamma_s^2}{(1-\mathcal{N}_0)^2\ln(1-\mathcal{N}_0)}, \qquad \mathcal{B} = \frac{1}{2}\left(1-\mathcal{N}_0\right)^{-\frac{(1-\mathcal{N}_0)}{\mathcal{N}_0\gamma_s}}.
\end{eqnarray}
The free parameters of the  CGC dipole model, $\sigma_0,~\lambda,~ x_0$, and $\gamma_s$ are determined by a fit to the H1 and ZEUS (2015) $F_2$ structure function data~\cite{H1:2015ubc} (for $x_{\rm Bj} \leq 0.01$ and $Q^2 \in [0.045,45]$ GeV$^{2}$) with a $\chi^2/\mbox{d.o.f}=1.03$~\cite{PhysRevD.94.074018}. Here, we use the %\sout{same CGC dipole model} 
parameters as determined in Ref.~\cite{PhysRevD.94.074018} as: %\sout{and their numerical values are}
$\sigma_0 = 26.3$ mb, $\gamma_s = 0.741,  \lambda = 0.219$, and $x_0 = 1.81 \times10^{-5}$ for $m_{u,d}\sim 0.046$ GeV. {The parameters} $\mathcal{N}_0$ and $\kappa_0$ are fixed as $0.7$ and $9.9$ (leading order
Balitsky-Fadin-Kuraev-Lipatov prediction), respectively.

To compute the scattering amplitude for exclusive $\rho$-meson production,  Eq.~(\ref{eq:amplitude-VMP}), we need to employ the LFWFs of the incoming virtual photon and the outgoing vector meson. In practice, the expressions for the photon LFWFs are obtained perturbatively in light-front QED. The lowest order perturbative  LFWFs for the longitudinally  and transversely  polarized photons are given by~\cite{PhysRevD.22.2157,PhysRevD.55.2602,Kulzinger:1998hw},
\begin{eqnarray}\label{eq:photonLFWFs}\nonumber
\Psi^{\gamma,\Lambda=0}_{h,\bar{h}}(x,{\mathbf r}_{\perp}; Q^2, m_q)  &=& \sqrt{\frac{N_{c}}{4\pi}}\delta_{h,-\bar{h}}e\, e_{q}2 x(1-x) Q \frac{K_{0}(\epsilon {\mathbf r}_{\perp})}{2\pi}, 
\label{photonwfL} \\
\Psi^{\gamma,\Lambda=\pm 1}_{h,\bar{h}}(x,{\mathbf r}_{\perp}; Q^2, m_q) &=& \pm \sqrt{\frac{N_{c}}{2\pi}} e \, e_{q} 
 \big[i e^{ \pm i\theta_{{\mathbf r}_{\perp}}} (x \delta_{h\pm,\bar{h}\mp} -  (1-x) \delta_{h\mp,\bar{h}\pm}) \partial_{{\mathbf r}_{\perp}}   +  m_{q} \delta_{h\pm,\bar{h}\pm} \big]\frac{K_{0}(\epsilon {\mathbf r}_{\perp})}{2\pi},
\end{eqnarray}
where $\epsilon^2=x(1-x)Q^{2} + m_{q}^{2}$ and $e^2=4\pi\alpha_{\rm em}$ with $\alpha_{\rm em}$ being the QED coupling constant, $e_{q}$ and $m_q$ represent the effective charge and mass of the quark, respectively. $N_{c}$ corresponds to the color factor, $K_{0}$ denotes the second kind of Bessel function and $r_{\perp}e^{ i\theta_{{\mathbf r}_{\perp}}}$ is the complex
notation for the transverse distance between the quark
and the antiquark. On the other hand, a nonperturbative
model for the meson LFWFs is discussed in Sec.~\ref{sec:HLFQCD}.

The total cross section is expressed {as the linear combination} of the transverse and longitudinal cross sections {by integrating them (given in Eq.~\eqref{eq:diffcrosssect}) over $t$. Therefore,}% \sout{and it is obtained by integrating Eq.~(\ref{eq:diffcrosssect}) over $t$,}
\begin{eqnarray}
	\sigma_{\textrm{tot}}^{\gamma^{\ast}p\rightarrow Vp}(x,Q^{2})=\sigma_{\Lambda=\pm 1}^{\gamma^{\ast}p\rightarrow Vp}(x,Q^{2})+\omega \sigma_{\Lambda=0}^{\gamma^{\ast}p\rightarrow Vp}(x,Q^{2}),
\end{eqnarray}
where $\omega$ is the photon polarization parameter, with $\langle\omega\rangle=0.98$ in the kinematic domain corresponding to the HERA measurement for the $\rho$-meson production~\cite{H1:2009cml}. We consider the same value of $\omega$ for predicting the total cross section and compare it with the HERA data.
% 

%%%%%%%%%%%%%%%%%
\section{HOLOGRAPHIC MESON WAVE FUNCTIONS with LONGITUDINAL CONFINEMENT}\label{sec:HLFQCD}
%%%%%%%%%%%%%%%%%
 In the previous section, we reviewed the photon wave functions, which are obtained from the perturbative QED. The vector meson LFWFs appearing in Eq.~(\ref{eq:amplitude-VMP}) can not be obtained in perturbation theory. %{\color{red}????}. 
 Nevertheless, they can be considered to have the same spinor and polarization structure as in the photon case, together with  an unidentified nonperturbative wave function. 
Various ansatz for the meson nonperturbative wave functions have been reported in literature~\cite{PhysRevD.55.2602,Nemchik:1996cw,PhysRevLett.109.081601}. However, the most popular is the boosted Gaussian wave function~\cite{Nemchik:1996cw, PhysRevD.69.094013}, which has recently been employed in Refs.~\cite{PhysRevD.69.074016,PhysRevD.87.034002} to simultaneously reproduce the cross-section data for diffractive $\rho, \phi$ and $J/\Psi$ production.
 Explicitly, the spin-improved LFWFs for longitudinally and transversely polarized vector meson can be written as~\cite{PhysRevD.69.094013,PhysRevLett.109.081601,Kaur:2020emh, PhysRevD.102.034021}
\begin{eqnarray}
\Psi^{V,\Lambda=0}_{h,\bar{h}}(x,{\mathbf r}_{\perp}) =  \frac{1}{2} \delta_{h,-\bar{h}}  \bigg[ 1 + 
{ m_{q}^{2} -  \nabla_{{\mathbf r}_{\perp}}^{2}  \over x(1-x)M^2_{V} } \bigg] \Psi(x,{\mathbf r}_{\perp}) \,,
\label{eq:LFWFlong}
\end{eqnarray}
and
\begin{eqnarray}
\Psi^{V, \Lambda=\pm 1}_{h,\bar{h}}(x,{\mathbf r}_{\perp}) = \pm \bigg[  i e^{\pm i\theta_{{\mathbf r}_{\perp}}}  ( x \delta_{h\pm,\bar{h}\mp} - (1-x)  \delta_{h\mp,\bar{h}\pm})  \partial_{{\mathbf r}_{\perp}}+ m_{q}\delta_{h\pm,\bar{h}\pm} \bigg] {\Psi(x,{\mathbf r}_{\perp}) \over 2 x (1-x)}\,, 
\label{eq:LFWFtrans}
\end{eqnarray}
%{\color{red} Shouldn't r be bold or shown with the perp sign in the subscript of r, while writing as the indices of $\Psi$???}
respectively, where $\Psi$ represents the spin independent part of the vector meson wave functions. 

Brodsky and de T\'eramond proposed a nonperturbative approach to construct the hadronic LFWFs based on hLFQCD~\cite{PhysRevLett.94.201601,PhysRevLett.96.201601,PhysRevLett.102.081601}. 
To connect with AdS space, a holographic variable,    
$\zeta=\sqrt{x(1-x)}r_{\perp}$
is introduced, and the wave function is written in a factorized form in {terms of} $x$, $\zeta$ and $\varphi$ variables :   
\begin{equation}
	\Psi (x, \zeta, \varphi)= \frac{\phi (\zeta)}{\sqrt{2\pi \zeta}} e^{i L \varphi} X(x) \;,
	\label{full-mesonwf}
\end{equation}
where $X(x)=\sqrt{x(1-x)} \chi(x)$ and $\phi (\zeta)$ are referred as the longitudinal and transverse modes, respectively and $L$ is the orbital quantum number. 
In hLFQCD, only the transverse mode, $\phi(\zeta)$,  is dynamical and it is generated by the holographic  Schr\"{o}dinger-like equation~\cite{Brodsky:2006uqa,deTeramond:2005su,deTeramond:2008ht,Brodsky:2014yha}, 
\begin{equation}
	\left(-\frac {\mathrm{d}^2}{\mathrm{d} \zeta^2}+\frac{4L^2-1}{4 \zeta^2}+U_\perp(\zeta)\right)\phi(\zeta)= M_\perp^2 \phi(\zeta) \;,
	\label{SEq}
\end{equation}
where the confinement potential is given by
\begin{equation}
	U_\perp(\zeta)=\kappa^4 \zeta^2 + 2\kappa^2(J-1) \;,
	\label{U-LFH}
\end{equation}
with $J=L+S$ being the total angular momentum of the meson. The analytical expression of Eq.~\eqref{U-LFH} is uniquely determined by a holographic mapping to $\mathrm{AdS}_5$, where light-front variable $\zeta$ maps onto the fifth dimension of AdS space and the underlying conformal symmetry~\cite{Brodsky:2013ar}. The emerging mass scale, $\kappa$, fixes the confinement scale and produces meson masses in the chiral limit. Using Eq.~\eqref{U-LFH} {in} Eq.~(\ref{SEq}) {yields}
\begin{equation}
	M_{\perp}^2(n_\perp , J, L)=4\kappa^2\left(n_\perp + \frac{J+L}{2}\right) \;,
	\label{MTM}
\end{equation}
%\sout{with $n_\perp$ being the transverse principle quantum number,} 
and 
\begin{equation}
	\phi_{n_\perp L}(\zeta )\propto \zeta^{1/2+L}\exp\left(\frac{-\kappa^2\zeta^2}{2}\right)L_{n_\perp}^L(\kappa^2\zeta^2)\; ,
	\label{TMWF}
\end{equation}
with $n_\perp$ being the transverse principle quantum number.
An important outcome of Eq.~\eqref{MTM} is that the lowest-lying hadronic bound state, with $n_\perp =L=S=0$, is massless. This is inherently recognized as the pion, which is anticipated to exhibit zero mass in the chiral limit of QCD.

Meanwhile, the longitudinal mode, $\chi(x)$, is not dynamical in hLFQCD. The longitudinal wave function,  $X(x)=\sqrt{x(1-x)}$ is explicitly obtained by the holographic mapping of the electromagnetic or gravitational form factor in physical spacetime and $\mathrm{AdS}_5$~\cite{Brodsky:2007hb,Brodsky:2008pf}. This results $\chi(x)=1$. Inserting Eq.~\eqref{TMWF} in Eq.~\eqref{full-mesonwf} yields the holographic LFWFs in the chiral-limit. For the ground state mesons  ($n=0,\,L=0$), the holographic LFWFs become
\begin{equation}
	\Psi_{m_q=0} (x,\zeta^2) \propto \sqrt{x(1-x)} \exp\left(\frac{-\kappa^2\zeta^2}{2}\right)\;.
	\label{pichiralwf} 
\end{equation}
A two-dimensional Fourier transform results 
\begin{align}
	\Phi_{m_q=0}(x, k^2_\perp )  &\propto \frac{1}{\sqrt{x(1-x)}}\exp\left(-\frac{k_\perp^2}{2\kappa^2 x(1-x)}\right)\,,
		\label{FT}
\end{align}
%{\color{red}perp in subscript of r everywhere, to be consistent with the notation of k?}
where  ${k}_\perp=|\mathbf{k}_\perp|$, the Fourier conjugate of $\mathbf{r}_{\perp}$, defines the transverse momentum of the quark. 
Going beyond the chiral limit, Brodsky and de T\'eramond proposed a prescription {to describe the longitudinal mode as}%\sout{such thatthe longitudinal mode can be written as}
~\cite{Brodsky:2008pg}:
\begin{align}
	X(x) &= \sqrt{x(1-x)}\exp\left[-\frac{1}{2\kappa^2}\left(\frac{m_q^2}{x}+\frac{m_{\bar{q}}^2}{1-x} \right)\right]\;,
	\label{bda}
\end{align}
based on the observation that the chiral-limit of invariant mass of quark-antiquark pair,
\begin{equation}
	{\cal M}^2_{q\bar q}=\frac{k_\perp^2}{x(1-x)}+\frac{m_q^2}{x}+\frac{m_{\bar{q}}^2}{1-x}\; ,
\end{equation}
appears in Eq.~\eqref{FT}. Consequently, the bound-state mass eigenvalue  receives a first-order correction such that
\begin{equation}
	\Delta M^2 = \int \frac{\mathrm{d} x}{x(1-x)} X^2(x) \left(\frac{m_q^2}{x}+\frac{m_{\bar{q}}^2}{1-x} \right)\;.
	\label{massshift}	
	\end{equation}
Note that there are two shortcomings with the above prescription. First, it indicates that~\cite{Li:2021jqb}
$M_{\pi}^2=\Delta M^2 \propto 2m_q^2 (\ln(\kappa^2/m_q^2)-\gamma_E)$
with $\gamma_E=0.577216$ being the Euler's constant, in contrast to Gell-Mann-Oakes-Renner (GMOR) relation, $M_{\pi}^2\propto m_q$.  Second, the longitudinal mode, given by  Eq. \eqref{bda}, with no nodes, {remains same for all the radially excited states}. %\sout{is also assumed for all radially excited states}. 
However, this  prescription {has successfully been implemented to describe the light as well as heavy mesons} %\sout{has been widely used in a successful phenomenology of light mesons as well as heavy mesons}
~\cite{Brodsky:2014yha,Brodsky:2008pg,deTeramond:2021yyi,Swarnkar:2015osa,Ahmady:2020mht,Ahmady:2019hag,Ahmady:2019yvo,Ahmady:2018muv,Ahmady:2016ufq}.

Meanwhile, Refs.~\cite{Ahmady:2021lsh,Ahmady:2021yzh} consider longitudinal dynamics, generated by the 't Hooft Equation, in order to describe the full meson spectrum, while the pion  dynamics has been predicted in Ref.~\cite{Ahmady:2022dfv}. The concept of employing the 't Hooft equation to extend beyond the invariant mass prescription was initially suggested in Ref.~\cite{Chabysheva:2012fe}, aiming to forecast meson decay constants and  parton distribution
functions.
%{\color{red} also PDFs???}. 
Recently, in Refs.~\cite{deTeramond:2021yyi,Li:2021jqb}, the prescription was surpassed using a phenomenological longitudinal confinement potential, which was initially introduced in Ref.~\cite{Li:2015zda} within the framework of basis light-front quantization. While both Refs.~\cite{deTeramond:2021yyi,Li:2021jqb} concentrate on the chiral limit and the occurrence of chiral symmetry breaking, Ref.~\cite{deTeramond:2021yyi} broadens their investigation to heavy mesons in their ground state and explores the connection of their approach to the 't Hooft equation. It is worth mentioning that there has been a notable surge in interest regarding the incorporation of longitudinal dynamics within hLFQCD~\cite{Li:2021jqb,deTeramond:2021yyi,Lyubovitskij:2022rod,Weller:2021wog,Ahmady:2021lsh,Rinaldi:2022dyh}.

The 't Hooft equation can be derived by using the QCD Lagrangian in (1+1)-dim with large $N_{c}$ approximations as~\cite{tHooft:1974pnl}
\begin{eqnarray}
\left(\frac{m_{q}^{2}}{x}+\frac{m_{\bar{q}}^{2}}{1-x}\right)\chi(x)+\frac{g^{2}}{\pi}\mathcal{P}\int {\mathrm d}y \frac{|\chi(x)-\chi(y)|}{(x-y)^{2}}=M_{\parallel}\chi(x)\,,
\label{eq:tHooft}
\end{eqnarray}
where $g$ is the longitudinal  confinement scale and $\mathcal{P}$ denotes the Cauchy principal value. It is important to note that in the conformal limit, the 't Hooft equation has a gravity dual on AdS$_{3}$~\cite{Katz:2007br} and has been widely studied in the literature~\cite{Zhitnitsky:1985um,PhysRevD.57.1366,PhysRevLett.69.1018,PhysRevD.103.074002,PhysRevD.62.094011,PhysRevD.104.036004,Ahmady:2021lsh,Ahmady:2022dfv,PhysRevD.104.074013}. Unlike the holographic light-front Schr\"odinger equation, the 't Hooft does not admit analytical solutions. We solve it numerically using the matrix method illustrated in Ref.~\cite{Chabysheva:2012fe}. 
Using both the holographic Schr\"odinger equation and the 't Hooft equation, the meson mass is then given by 
\begin{align}
	M^2(n_\perp ,n_\parallel ,J, L)= 4\kappa^2\left(n_\perp + \frac{J+L}{2}\right)+ M_\parallel^2(n_\parallel , m_q, m_{\bar q}, g)\; ,
	\label{totalmass}
\end{align}
where $n_\parallel$ defines the longitudinal quantum number. Since, the holographic Schrödinger equation
predicts a massless pion, it follows that the only contribution to the pion mass is produced by the ’t Hooft equation. Note that together, the holographic Schrödinger equation and the ’t Hooft equation correctly predicts the GMOR relation $M_{\pi}^{2}\sim m_{u,d}$~\cite{deTeramond:2021yyi,Ahmady:2022dfv}. With the universal transverse confinement scale $\kappa=0.523$ GeV and the light quark mass $m_{u/d}=0.046$ GeV, which is the value considered in hLFQCD together with the IMA~\cite{Brodsky:2014yha}, the longitudinal confining scale $g=0.109$ GeV leads to excellent agreement of the mass spectroscopy for the pion family with the experimental data~\cite{Ahmady:2022dfv}. In this work, we use the same set of parameters %\sout{, i.e., $m_{u/d}=0.046$ GeV, $\kappa=0.523$ GeV, and $g=0.109$ GeV,} 
in order to predict the spectroscopic data for the $\rho$-meson family and obtain the corresponding wave functions.

The complete spin-independent part of the meson LFWFs can then be expressed as,
\begin{eqnarray}
\Psi (x,\zeta) = \mathcal{N} \sqrt{x (1-x)}
\chi(x)\exp{ \left[ -{ \kappa^2 \zeta^2  \over 2} \right] } ,
\label{eq:totalhwf}
\end{eqnarray}
where $\chi(x)$ is the solution of the ’t Hooft equation.  $\mathcal{N}$ is a  normalization constant, which depends on polarization of the meson and can be fixed by using the normalization condition as,
\begin{eqnarray}
\sum_{h,\bar{h}} \int {\mathrm d}^2 {\mathbf{r}_{\perp}} \, {\mathrm d} x |
\Psi^{V, \Lambda} _{h, {\bar h}}(x, {\mathbf r}_{\perp})|^{2} = 1 ,
\label{eq:normalization}
\end{eqnarray}
where the forms of the {spin-improved} %\sout{full} 
wave functions $\Psi_{h,\bar{h}}^{V,\Lambda}$ are given in Eqs.~(\ref{eq:LFWFlong}) and (\ref{eq:LFWFtrans}). 
The numerical solutions for the longitudinal modes $\chi(x)$ of the ground state meson can  approximately be fitted to the following polynomial form: 
\begin{eqnarray}
\chi(x)\simeq x^{\beta_{1}}(1-x)^{\beta_{2}},
\label{eq:tHooftanaly}
\end{eqnarray}
with $\beta_i$ being the quark mass dependent variables, which vanish in the chiral limit. For the ground state of $\rho$-meson, we find that $\beta_{1,2}=0.51$.
%

%========================
\section{Results and discussion}\label{sec:results}
%========================
\begin{table}
\caption{Quantum numbers and masses of the $\rho$-meson family {{with $S=1$}}.}
%\vspace{0.1cm}
\centering
\begin{tabular}{|c c c c c c|}
\hline
$J^{P(C)}$ & Name & $n_\perp$ & $n_\parallel$ & $L$ & $M$ (MeV) \\
\hline
$1^{--}$ & $\rho(770)$ & 0 & 0 & 0 & 752 \\
$2^{++}$ & $a_2(1320)$ & 0 & 2 & 1 & 1315 \\
$3^{--}$ & $\rho_3(1690)$ & 0 & 4 & 2 &  1702 \\
$4^{++}$ & $a_4(1970)$ & 0 & 6 & 3 & 2016 \\
$1^{--}$ & $\rho(1450)$ & 1 & 2 & 0 & 1315 \\
$2^{++}$ & $a_2(1700)$ & 1 & 4 & 1 & 1702 \\
$1^{--}$ & $\rho(1700)$ & 2 & 4 & 0 & 1702 \\
\hline
\end{tabular}
\label{Spectroscopy}
\end{table}
%========================
\subsection{Mass spectroscopy}
\begin{figure}[hbt!]
	\begin{center}
		\includegraphics[width=0.5\linewidth]{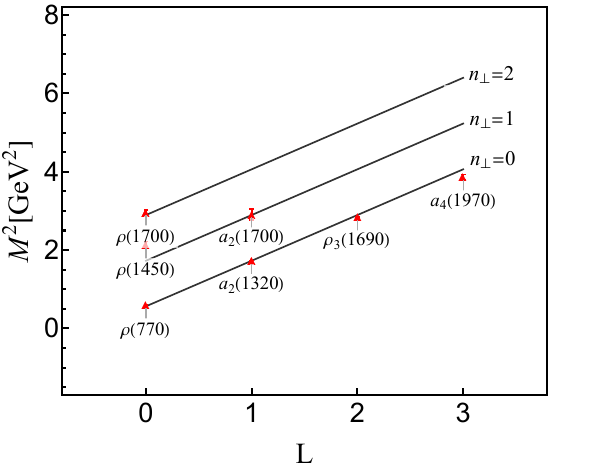}
		\caption{Our Regge trajectories for the $\rho$-meson family.}
		\label{pionmass}
	\end{center}
\end{figure}

%=====================
The  parity and charge conjugation quantum numbers of meson states are given by~\cite{Ahmady:2021lsh,Ahmady:2021yzh,Ahmady:2022dfv}:
\begin{equation}
	P=(-1)^{L+1}\; ,\;\; C=(-1)^{L+S+n_\parallel}\; .
	\label{pcrules}
\end{equation} 
Using Eq.~\eqref{totalmass} and the parameters determined for the pion family: $m_{u/d}=0.046$ GeV, $\kappa=0.523$ GeV, and $g=0.109$ GeV~\cite{Ahmady:2022dfv}, we are able to describe the spectroscopic data for the $\rho$-meson family. We present our computed masses of the {$\rho$-meson and its excited states} in Table~\ref{Spectroscopy}.  Our results (last column)  are in good agreement with the experimental data (second column, in parentheses). Note  that an emerging condition $n_\parallel\ge n_\perp +L$ in Table~\ref{Spectroscopy} is observed to remain true across the full hadron spectrum \cite{Ahmady:2021yzh}.  The resulting Regge trajectories for the {$\rho$-meson family} in our calculation are shown in Fig.~\ref{pionmass}. 
%=============================
%=============================

%=============================
\subsection{$\rho$-meson diffractive cross-section}
\begin{figure}
	\centering
	\includegraphics[scale=0.57]{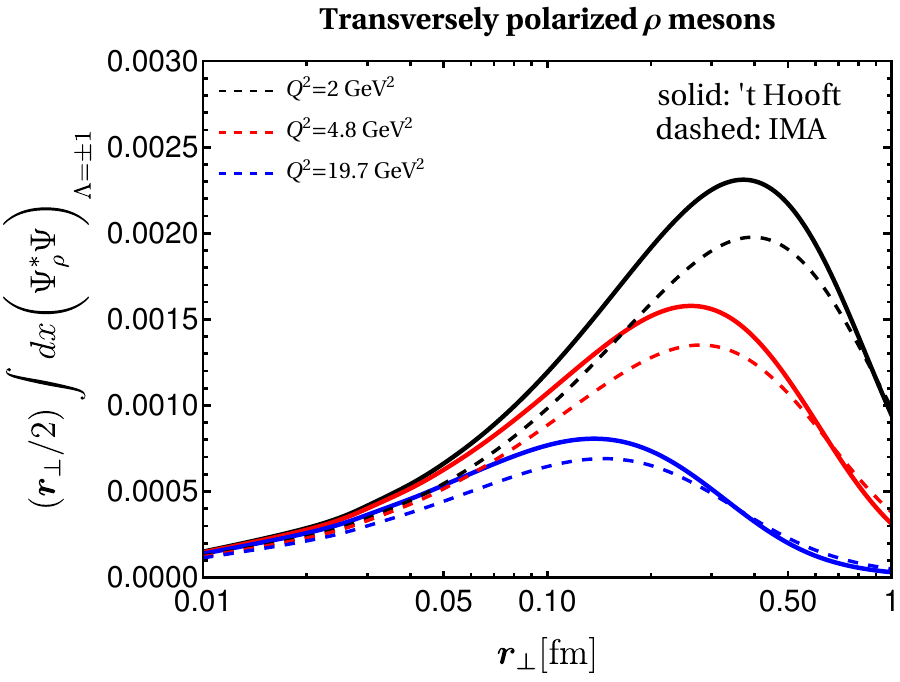}\hspace{0.2cm}
	\includegraphics[scale=0.57]{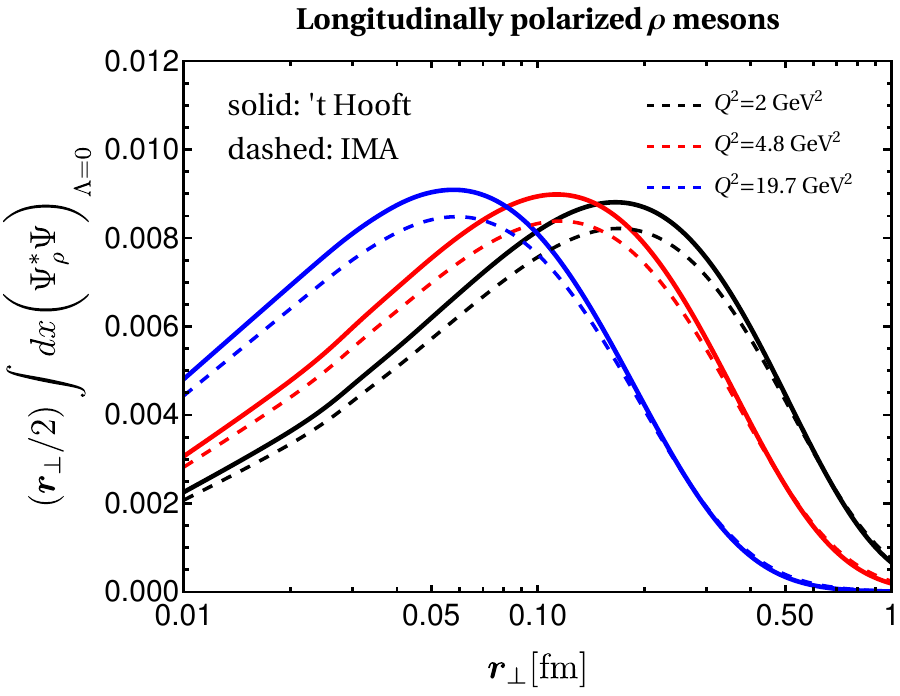}
	\caption{The transverse (left) and the longitudinal (right) overlap functions between the photon and the $\rho$-meson LFWFs integrated over $x$ as a function of the dipole transverse size ${\mathbf r}_{\perp}$ {(in fm)} at different photon virtualities predicted by the light front holography 't Hooft (solid) and the light front holography IMA (dashed) approaches, respectively.}
  \label{fig:overlaprep}
\end{figure}

The exclusive vector meson  production cross section depends upon the overlap of the $q\bar{q}$ component of the virtual photon wave functions with the vector meson wave functions as illustrated in Eq.~(\ref{eq:amplitude-VMP}). We present the overlap of the LFWFs after integrating over $x$ at different photon virtualities $Q^{2}=2,\, 4.8$ and $19.7$ GeV$^{2}$ in Fig.~\ref{fig:overlaprep}. We  compare two different overlap functions, which are different in considering the longitudinal modes in the meson wave functions: (i) the IMA that does not contain dynamical mode along the longitudinal direction, and (ii) the one containing  longitudinal dynamics incorporated through the 't Hooft equation.  
However, we find that they lead to more or less similar behavior of the overlap functions. The peaks of the distributions undergo a shift towards lower values of {${\mathbf r}_{\perp}$} and decrease in magnitude as the virtuality of the photon increases.

\begin{figure}
	\centering
	\includegraphics[scale=0.62]{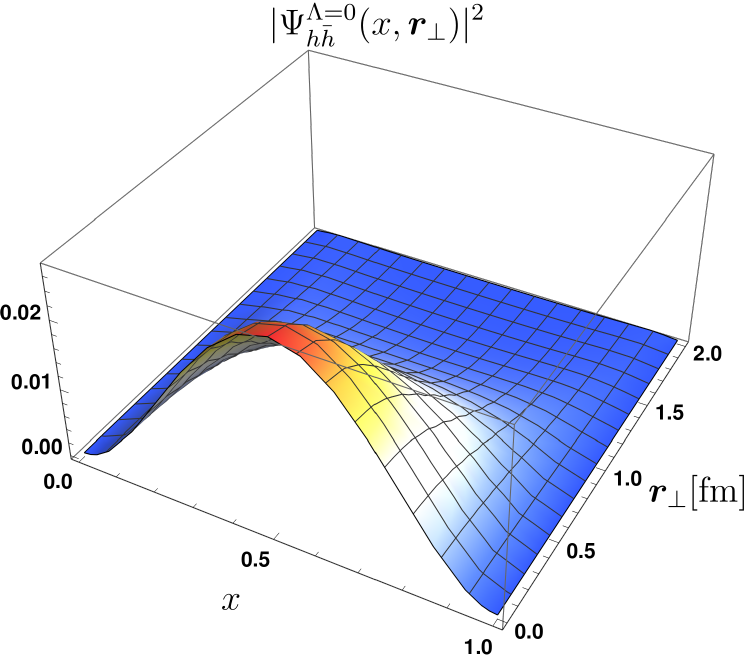}\hspace{0.8cm}
	\includegraphics[scale=0.62]{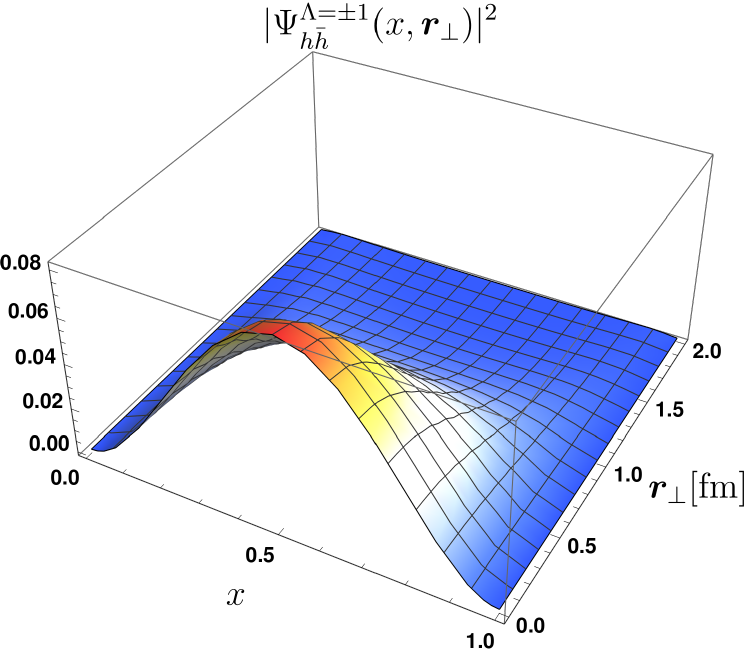}%\vspace{0.5cm}
	\caption{ Distributions of the longitudinal $|\Psi_{h,\bar{h}}^{\Lambda=0} (x,{\mathbf r}_{\perp})|^2$ (left) and transverse $|\Psi_{h,\bar{h}}^{\Lambda=\pm1} (x,{\mathbf r}_{\perp})|^2$ (right) LFWFs of the $\rho$-meson as functions of longitudinal momentum fraction carried by the quark $x$ and the dipole separation {$\mathbf{r}_{\perp}$} (in fm). }%\sout{{\color{red} Also write the units in legends and in captions, to be consistent}}}
	\label{fig:3Doverlaprep}
\end{figure}

\begin{figure}
	\centering
	\includegraphics[scale=0.68]{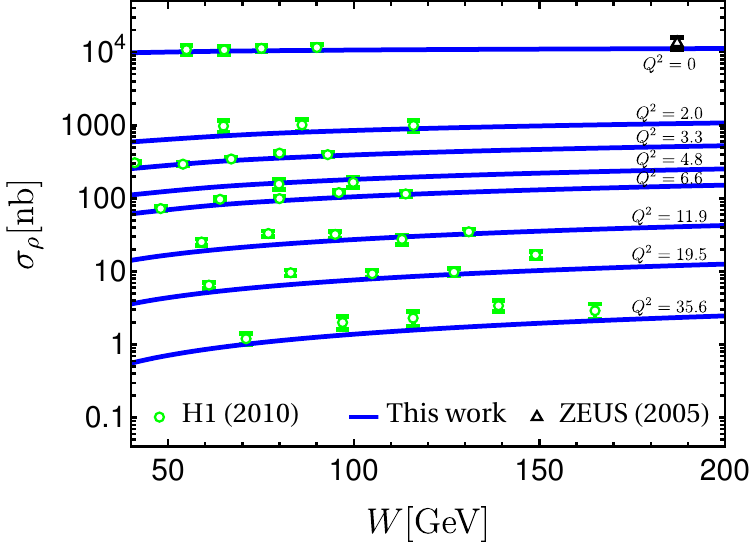}\hspace*{0.5cm}
	\includegraphics[scale=0.68]{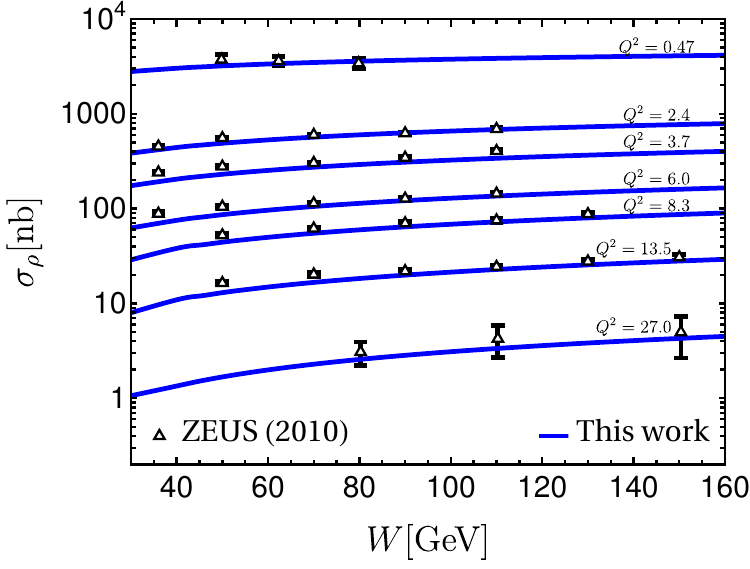}
	\caption{Our predictions of the total diffractive cross-section for $\gamma^{\ast}p\rightarrow \rho p$  as a function of $W$ {(in GeV)} in different $Q^{2}$ bins {(in GeV$^2$)} and compared with experimental data from H1 2010~\cite{H1:1996prv,H1:2009cml}, ZEUS 2000  data (at $Q^{2}=0$)~\cite{ZEUS:1997rof} (left) and ZEUS 2010 data
		~\cite{H1:2009cml} (right).}
	\label{fig:cross-sec_w_H1}
\end{figure}

\begin{figure}
	\centering	
	\includegraphics[scale=0.68]{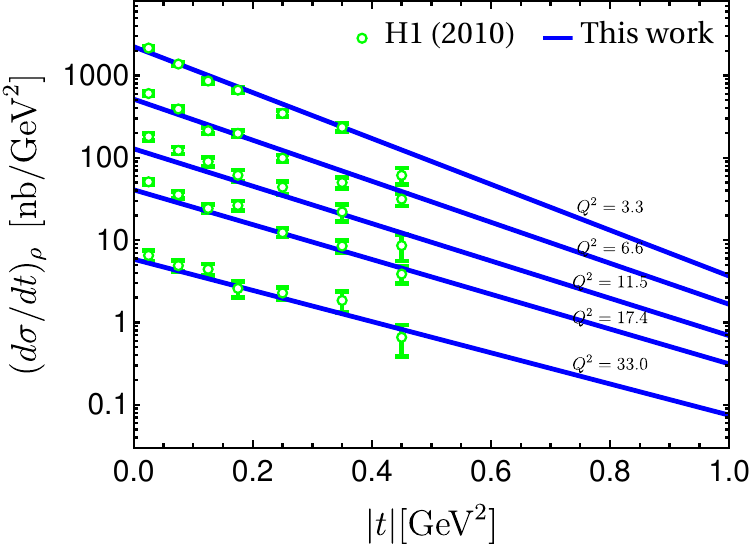}\hspace*{0.5cm}
	\includegraphics[scale=0.68]{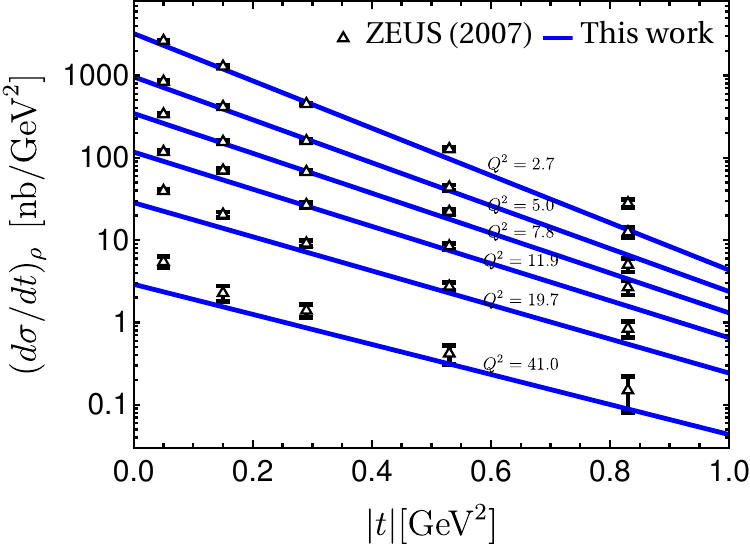}
	\caption{Our predictions of differential cross-section, $d\sigma/dt$ {(in nb/GeV$^2$)} for $\gamma^{\ast}p \rightarrow \rho^{0}p$ as a function of $|t|$ {(in GeV$^2$)} compared with %experimental data from 
		H1 2010~\cite{H1:2009cml} (left) and ZEUS 2007~\cite{ZEUS:2007iet} (right) data, respectively.}
	\label{fig:diff-cross-sec}
\end{figure}

In Fig.~\ref{fig:3Doverlaprep}, we illustrate the three-dimensional probabilistic distributions, $|\Psi_{h,\bar{h}}^\Lambda (x,{\mathbf r}_{\perp})|^2$, as a function of $x$ and ${\mathbf r}_{\perp}$ for a longitudinally (left panel) and a transversely (right panel) polarized $\rho$-meson from our resulting holographic LFWFs incorporated with longitudinal modes generated by the 't Hooft equation. We notice that the wave function peaks at $x = 0.5$ and ${\mathbf r}_{\perp}=0$, and go rapidly to zero as $x\rightarrow 0,1$ and ${\mathbf r}_{\perp}$ increases. %{\color{red}which shows that the contribution of large dipoles are suppressed.} 
Our results {behaves} similar to the wave functions reported 
%Similar kind of wave function behavior has also been reported 
in Ref.~\cite{PhysRevD.69.094013}.\\
We have now all the ingredients (photon and vector meson LFWFs, and the dipole cross-section) to compute the diffractive vector meson production cross section. We calculate $\rho$-meson production utilizing the CGC dipole model with parameters fitted to DIS data from HERA, as detailed in Ref.~\cite{PhysRevD.94.074018}.
% Note that all ourpredictions are generated using the same holographic wave function given by Eq.~(\ref{eq:totalhwf}). 
In Fig.~\ref{fig:cross-sec_w_H1}, we show the total cross section as a function of $W$ for different $Q^{2}$ bins. On the left panel, we compare our  predictions with the experimental data for $0\leq Q^{2}\leq35.6$ GeV$^{2}$ from the H1  Collaboration~\cite{H1:1996prv,H1:2009cml}, whereas, the right panel compares our results with the experimental data from the ZEUS Collaboration~\cite{H1:2009cml} in the range $0.47\leq Q^{2}\leq 27.0$ GeV$^{2}$. From the comparison, we observe that our predictions are in reasonable agreement with the experiments within the range of allowed uncertainty. We also note that the longitudinal dynamics implemented through the 't Hooft equation improves the predictions compared to those calculated using IMA in Ref.~\cite{PhysRevD.94.074018}. %as well as b-Sat and b-CGC models~\cite{PhysRevD.78.014016}
The differential cross-section, $d\sigma/dt$, as a function of $t$ for the elastic $\rho$-meson production are shown in Fig.~\ref{fig:diff-cross-sec}, where we compare our results with the  experimental data from the H1~\cite{H1:2009cml} (left panel) and ZEUS~\cite{ZEUS:2007iet} (right panel) Collaboration at different values of $Q^{2}$. Again, we observe that our predictions show a good agreement
with the measurements. %Whereas, we present the comparison of the differential cross-section for elastic $\rho$-meson production with ZEUS 2007 data~\cite{ZEUS:2007iet} (on right panel). 
{However, at large $t$ and small $Q^2$, our results are somewhat underestimated for the ZEUS data.}

\begin{figure}
	\centering	
\includegraphics[scale=0.68]{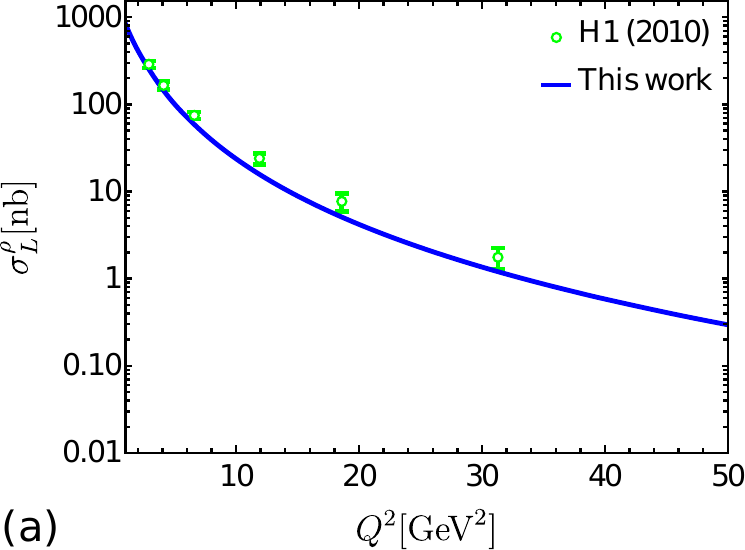}\hspace{0.5cm}
\includegraphics[scale=0.68]{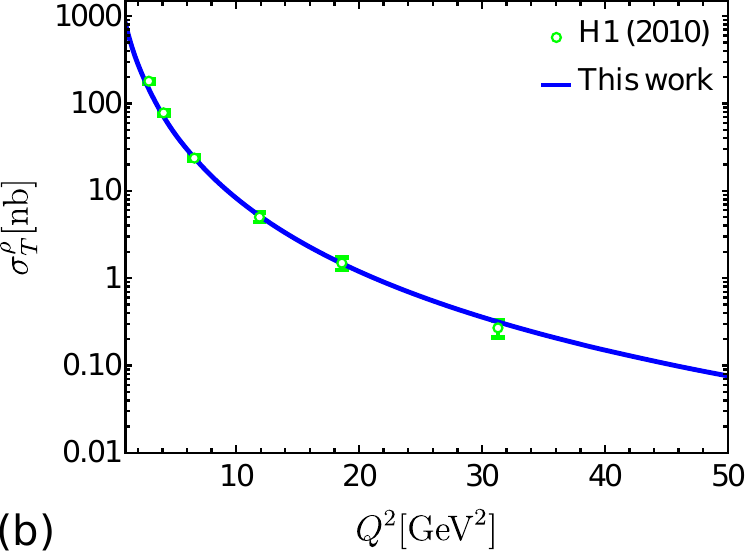}\vspace{0.5cm}
\includegraphics[scale=0.68]{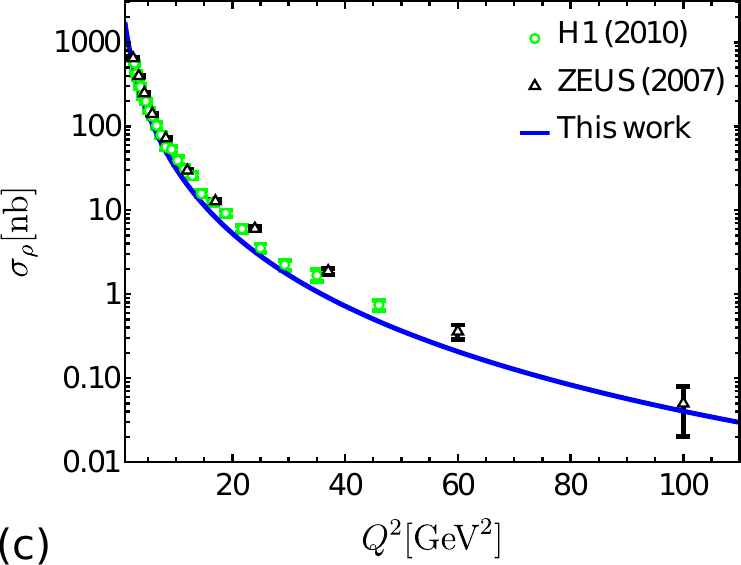}\hspace{0.5cm}
\includegraphics[scale=0.68]{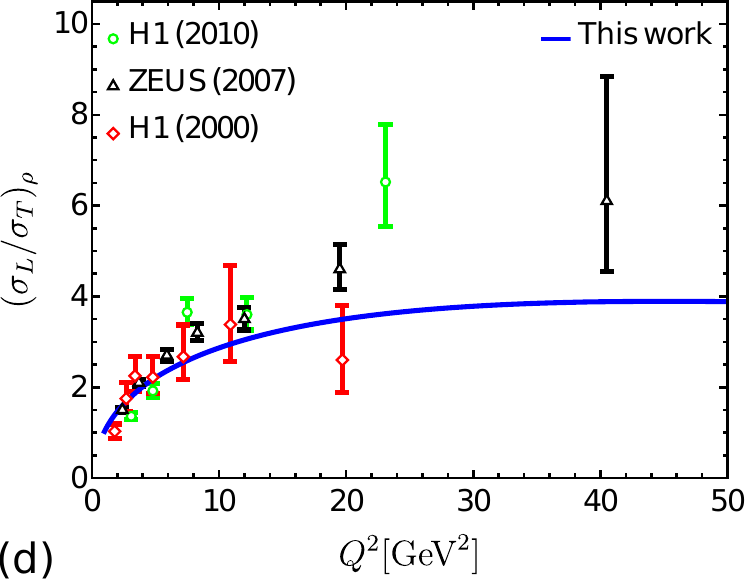}	
	\caption{Our predictions for the $Q^{2}$ dependence of (a) longitudinal, (b) transverse, (c) Total $\gamma^{\ast}p$ cross sections for elastic $\rho$ meson production with $W = 75$ GeV and (d) longitudinal to transverse cross-section ratio compared to the H1 at $W=75$ GeV~\cite{H1:1999pji,H1:2009cml} and ZEUS at $W=75$ GeV~\cite{ZEUS:2007iet} data.}
	\label{fig:total-cross-sec}
\end{figure}

The Fig.~\ref{fig:total-cross-sec}(a) and \ref{fig:total-cross-sec}(b) present the longitudinal and transverse cross-sections, respectively, for the diffractive $\rho$-meson production as a function of photon virtually for the %photon-proton center of mass energy 
{fixed value of} $W=75$ GeV. We find good agreement with the experiment at HERA~\cite{H1:2009cml}.
%In Fig.~\ref{fig:total-cross-sec} (b) we shows the transverse cross-section as a function of $Q^{2}$ for the same c.m. energy.
The total $\gamma^{\ast} p$ elastic cross-section for $\rho$-meson production is also found to be in good agreement  with the experiments~\cite{H1:1999pji,ZEUS:2007iet} as can be seen in  Fig.~\ref{fig:total-cross-sec}(c). Finally, in Fig.~\ref{fig:total-cross-sec}(d), we illustrate the ratio of the longitudinal to transverse cross section, $\sigma_{L}/\sigma_{T}$, for $\rho$ production. We compare our prediction with the H1 2010~\cite{H1:2009cml}, ZEUS 2007~\cite{ZEUS:2007iet} and old 2000 H1~\cite{H1:1999pji} data. Here, we notice that for the low values of $Q^{2}$, i.e., $Q^{2}\leq 10$ GeV$^{2}$, our results show a good agreement with all the three data sets, whereas it slightly deviates at large $Q^{2}$ from the measured data.
%
%%%%%%%%%%%%%%%%%%%%%%%%%%%%%%%%%
\subsection{Decay Constants and Distribution Amplitudes} \label{sec:DAs}
%%%%%%%%%%%%%%%%%%%%%%%%%%%%%%%%%
In this section, we compute the decay constants and distribution amplitudes for the $\rho$-meson and compare them with the experimental data along with the theoretical predictions {from other models.} %Note that we have determined the holographic wave function for the vector mesons~(\ref{eq:totalhwf}). 
The vector and tensor coupling constants, $f_V$ and $f_V^T$ respectively, are defined as the local vacuum-to-hadron matrix elements~\cite{PhysRevD.91.014018}
\begin{eqnarray}
	\langle 0|\bar q(0)  \gamma^\mu q(0)|V
	(P,\Lambda)\rangle =f_V M_{V} \epsilon_\Lambda^{\mu}\,,
	\label{eq:fv}
\end{eqnarray}
and
\begin{eqnarray}
	\langle 0|\bar q(0) [\gamma^\mu,\gamma^\nu] q(0)|V (P,\Lambda)\rangle =2 f_V^{T}  (\epsilon^{\mu}_{\Lambda} P^{\nu} - \epsilon^{\nu}_{\Lambda} P^{\mu}) \;,
	\label{eq:fvT}
\end{eqnarray}
%where $f_{V}$ and $f_{V}^{T}$ are the vector and tensor coupling constants, $M_{\rho}$ is the $\rho$ vector meson mass, 
where $q(\bar{q})$ are the quark(anti-quark) field operators at same space-time points. The momentum and polarization vectors  are denoted as $P^{\mu}$ and $\epsilon_{\Lambda}^{\mu}$, respectively. 
%By inserting the Fock state expansion of the meson state into Eqs.~(\ref{eq:fv}) and (\ref{eq:fvT}), one can obtain the decay constant expressions in terms of LFWFs~\cite{PhysRevD.87.054013}
In terms of LFWFs, the decay constant can be expressed as~\cite{PhysRevD.87.054013}
\begin{eqnarray}
f_{V} &=&  {\sqrt \frac{N_c}{\pi} }  \int_0^1 {\mathrm d} x  \left[ 1 + { m_{q}^{2}-\nabla_{\mathbf{r}_{\perp}}^{2} \over x (1-x) M^{2}_{V} } \right] \left. \Psi(x,\mathbf{r}_{\perp}) \right|_{\mathbf{r}_{\perp}=0}\,,
\label{eq:fvL}	
\end{eqnarray}
and
\begin{eqnarray}
	f_{V}^{T}(\mu) =\sqrt{\frac{N_c}{2\pi}} m_q \int_0^1 {\mathrm d} x \; \int {\mathrm d} \mathbf{r}_{\perp} \; \mu J_1(\mu \mathbf{r}_{\perp})  \frac{\Psi(x,\mathbf{r}_{\perp})}{x(1-x)}
	\label{eq:fvT1}\,,
\end{eqnarray}
%
%\sout{{\color{red} again $r_\perp$ in above equations}}
where $\Psi$ is the meson %and $\Psi_{T}$ are the normalized longitudinal and transverse components of the 
LFWF %meson light-front wave functions 
 given in Eq.~(\ref{eq:totalhwf}) and % \sout{ $m_{f}$ is the flavor masses and }
$\mu$ is the ultraviolet cut-off scale. We note that 
%the dependence of $\mu$ is not much for the tensor coupling, i.e., 
our predictions for the tensor coupling are scale independent for $\mu^{2}\geq1$. However, it is  sensitive to the quark mass $m_{q}$, as  can be seen from Eq.~(\ref{eq:fvT1}). In the chiral limit, $m_{q}\rightarrow 0$, the tensor coupling  vanishes, whereas the vector coupling has a nonzero value. The vector coupling can be used to calculate the electronic decay width
\begin{eqnarray}
	 \Gamma_{V \rightarrow e^+ e^-}={ 4 \pi  \alpha_{em}^2  C_V^2 \over 3 M_V }f_V^2  \,,
	\label{eq:decaywidth}
\end{eqnarray}
where, for the $\rho$-meson, $C_{\rho}=1/\sqrt{2}$. In Table~\ref{tab:Decay-width}, we present our predictions for the vector coupling constant, which is associated with the decay width of the $\rho$-meson. These predictions are compared with the results from LFhQCD with IMA approach~\cite{PhysRevD.94.074018,PhysRevLett.109.081601} and experimental data %available from the Particle Data Group (PDG)
~\cite{ParticleDataGroup:2014cgo}. Additionally, in Table~\ref{tab:fv-rho}, we compare our model predictions for 
decay constants when $\rho$-meson is considered to be longitudinally and transversely polarized, as well as their ratio $f_{\rho}^{\perp}/f_{\rho}$. % , using light quark masses $m_{u,d}\simeq 0.046$. 
We compare our results with  the predictions from other theoretical approaches in Table~\ref{tab:fv-rho}.
%those obtained from other non-perturbative studies, as mentioned in the table caption. 
Our prediction for the vector coupling constant demonstrates good agreement with both the theoretical and experimental studies. However, the tensor coupling constant is significantly smaller as compared to the other predictions in the literature. %coupling constants.
\begin{table}
	\caption{Our predictions for the electronic decay widths of the $\rho$-meson {using 't Hooft equation} %\sout{using the holographic and 't Hooft wavefunction given by Eq. \eqref{eq:totalhwf}} 
with $m_{u,d}=0.046$ GeV. The LF holography IMA predictions correspond to $m_{u,d}= [0.046,0.14]$ GeV.}
	\centering
	\begin{tabular}{c|c|c}
		\hline\hline
		$\rho$-meson   &  $f_V$ [MeV] & $\Gamma_{e^+e^-}$ [KeV] \\ \hline
		This work & $208$ & $ 6.294 $  \\ \hline
		LFH (IMA)\cite{PhysRevD.94.074018,PhysRevLett.109.081601} & $[210,211]$ & $[6.355,6.66]$  \\ \hline
		Exp. (PDG)~\cite{ParticleDataGroup:2014cgo} & $216(5)$ & $7.04 \pm 0.06$ \\ \hline\hline
	\end{tabular}
	\label{tab:Decay-width}
\end{table}
\begin{table}
\caption{Using the value of light quark masses, $m_{u,d}=0.046$ GeV, we predict the longitudinal and transverse decay constants for the $\rho$-meson, as well as the ratio between them and compared with other model estimate. Our predictions are at a scale of $\mu=1$ GeV.}% 
  \centering
  \begin{tabular}{c|c|c|c|c}
    \hline\hline
    Reference & Approach & $f_{\rho}$ [MeV] & $f_{\rho}^{\perp}$ [MeV] & $f_{\rho}^{\perp}/f_{\rho}$\\ \hline
    This work & LFH ($`$t Hooft) & $208$  & $36$ & $0.17$\\ \hline
    Ref.~\cite{PhysRevD.94.074018,PhysRevD.88.074031} & LFH (IMA) & $[211,214]$  & $[95,36]$ & $[0.45,0.17]$\\ \hline
    Ref. \cite{PhysRevC.92.055203} & LFQM & $208 \pm 7$ & $152 \pm 9$ &   \\ \hline
    Ref. \cite{PhysRevD.58.094016} & Sum Rules & $198 \pm 7$ & $152 \pm 9$ &   \\ \hline
%    Ref. \cite{PhysRevD.58.094016} & Sum Rules & $198 \pm 7$ & $152 \pm 9$ &   \\ \hline
    Ref. \cite{Ball:2007zt} & Sum Rules & $206 \pm 7$ & $145 \pm 9$ &  $0.70 \pm 0.04$ \\ \hline
     Ref. \cite{Becirevic:2003pn} & Lattice (continuum) &  &  & $0.72 \pm 0.02$ \\ \hline
     Ref. \cite{PhysRevD.68.054501} & Lattice (finite) &  &  & $0.742 \pm 0.014$ \\ \hline
     Ref. \cite{PhysRevD.80.054510} & Lattice (unquenched) &  & $159 \pm 0.008$ & $0.76 \pm 0.04$ \\ \hline
    Ref. \cite{PhysRevD.90.014011} & Dyson-Schwinger  & $212$ & $156$ & $0.73$ \\ \hline
    Ref. \cite{PhysRevD.75.034019} & LFQM: Linear~[HO]  & $246~[215]$ & $187~[163]$ & $0.76~[0.80]$ \\ \hline\hline
\end{tabular}
  \label{tab:fv-rho}
\end{table}

PDAs are obtainable through the vacuum-to-meson transition matrix elements of quark-antiquark nonlocal gauge-invariant operators. 
The longitudinal and transverse components of the distribution amplitude for the vector meson are defined as~\cite{PhysRevD.54.2182} 
\begin{eqnarray}
	f_{\rho}\phi_{\rho}^{\|}(x,\mu)=\int {\mathrm d}x^{-}e^{ixP^{+}x^{-}}\langle 0|\,\bar{q}(0)\gamma^{+}q(x^{-})\,|V(P,\Lambda=0)\rangle \,,
	\label{eq:DALong}
\end{eqnarray}
and
\begin{eqnarray}
	f_{\rho}^{\perp}\phi_{\rho}^{\perp}(x,\mu)=\frac{1}{2}\int {\mathrm d}x^{-}e^{ixP^{+}x^{-}}\langle 0|\,\bar{q}(0)[\epsilon^{\ast}_{\Lambda(\pm)}.\gamma,\gamma^{+}]q(x^{-})\,|V(P,\Lambda=\pm 1)\rangle\,.
	\label{eq:DAtrans}
\end{eqnarray}
After calculating the matrix elements, the above Eqs.~(\ref{eq:DALong}) and (\ref{eq:DAtrans}) lead to
\begin{eqnarray}
	\phi_\rho^{\|}(x, \mu)=  \frac{N_c}{\pi f_\rho M_\rho} \int \mathrm{d} \mathbf{r}_{\perp} \mu J_1(\mu \mathbf{r}_{\perp})\left[M_V^2 x(1-x)\right.
	 \left.+m_q^2-\nabla_{\mathbf{r}_{\perp}}^2\right] \frac{\Psi(x,\mathbf{r}_{\perp})}{x(1-x)}\, ,
\end{eqnarray}
and
\begin{eqnarray}
	\phi_\rho^{\perp}(x, \mu)=\frac{N_c m_q}{\pi f_\rho^{\perp}} \int \mathrm{d} \mathbf{r}_{\perp} \mu J_1(\mu \mathbf{r}_{\perp}) \frac{\Psi(x,\mathbf{r}_{\perp})}{x(1-x)},
\end{eqnarray}
where $f_{\rho}$ and $f_{\rho}^{\perp}$ are the vector and tensor couplings, which are given in Eq.~(\ref{eq:fvL}) and (\ref{eq:fvT1}), respectively. The longitudinal ($\phi_{\rho}^{\|}$) and transverse ($\phi_{\rho}^{\perp}$) components of the PDAs can be normalized as~\cite{Choi:2007yu}
\begin{eqnarray}
	\int_{0}^{1}{\mathrm d}x \phi_{\rho}^{\|}(x,\mu)=1,~~~\textrm{and}~~~\int_{0}^{1}{\mathrm d}x \phi_{\rho}^{\perp}(x,\mu)=1\,.
\end{eqnarray}
%\sout{{\color{red}$\infty???$ in eq. 37}}
%
%
\begin{figure}
		\centering	
	\includegraphics[scale=0.68]{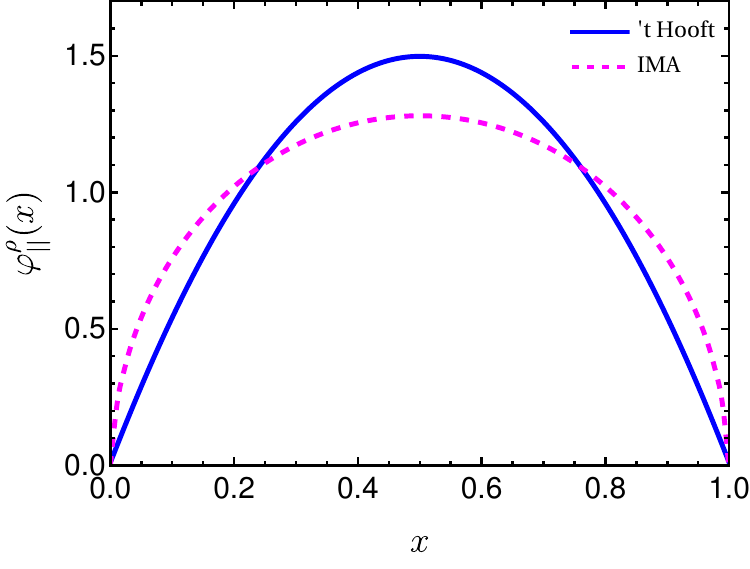}\hspace{0.5cm}
	\includegraphics[scale=0.68]{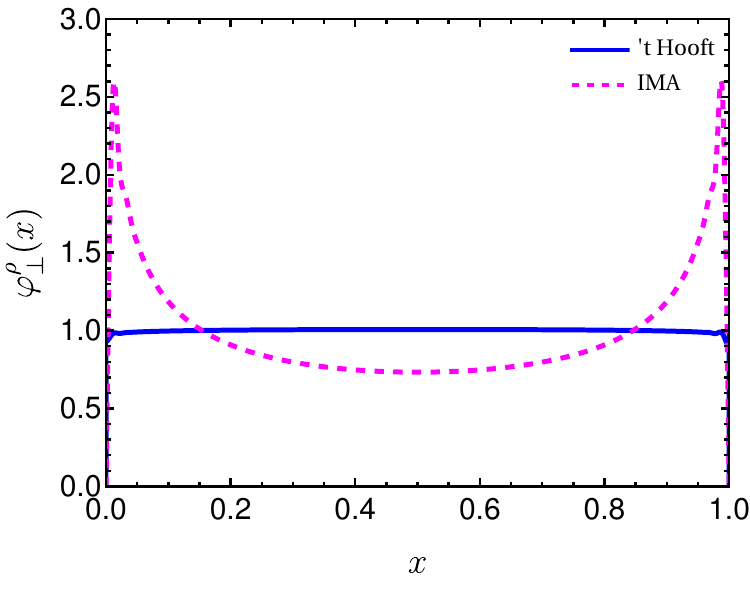}\vspace{0.5cm}
    \caption{Our results for the PDAs for longitudinally (left) and transversely (right) polarized $\rho$-mesons (in solid blue). We compared them with IMA  PDAs (magenta-dashed).}
	\label{Fig:DAs}
\end{figure}
Fig.~\ref{Fig:DAs} illustrates the normalized longitudinal and transverse components of $\rho$-meson PDAs and their comparison with hLFQCD associated with IMA predictions at nonzero light quark mass, $m_{q}\simeq 0.046$ GeV. We find that our $\phi_{\rho}^{\|}(x)$ exhibits narrower distribution, while $\phi_{\rho}^{\perp}(x)$ shows {flat} distribution compared to those evaluated using hLFQCD associated with IMA. 

We calculate the moments of the PDAs, also known as $\xi$-moments, in order to quantitatively compare with other approaches. The $n^{\rm th}$
moment is defined as~\cite{Choi:2007yu},
\begin{eqnarray}\label{zetam}
\langle\xi^{n}\rangle=	\langle (2x-1)^{n}\rangle=\int_{0}^{1}{\mathrm d}x(2x-1)^{n}\phi(x)\,.
\end{eqnarray}
In Table~\ref{Table:moments}, we compare our model results for the computed $\xi$-moments at $\mu \sim 1$ GeV with other theoretical estimations for $n=2,4,6,8~\textrm{and}~10$.  For the odd values of $n$, the moments vanish due to the isospin symmetry. Our $\xi$-moments are more or less consistent with other theoretical studies.

\begin{table}[t]
	\caption{Computed moments of the $\rho$-meson PDAs, Eq.~\eqref{zetam}, compared with selected results obtained elsewhere, using: AdS/QCD models for LFWFs fitted to HERA data \cite{Forshaw:2010py};
		QCD sum rules \cite{PhysRevD.54.2182,Ball:1998sk};
		nonlocal condensates \cite{Bakulev:1998pf,Pimikov:2013usa};
		light-front quark model \cite{PhysRevD.75.034019}; light-front holographic QCD~\cite{PhysRevD.94.074018},
		and lattice-QCD \cite{QCDSF-UKQCD:2007xbk,PhysRevD.83.074505}.
		\label{Table:moments}
	}
	\begin{tabular*}%{lcccc}
		{\hsize}
		{
			l@{\extracolsep{0ptplus1fil}}
			c|@{\extracolsep{0ptplus1fil}}
			l@{\extracolsep{0ptplus1fil}}
			l@{\extracolsep{0ptplus1fil}}
			l@{\extracolsep{0ptplus1fil}}
			l@{\extracolsep{0ptplus1fil}}
			l@{\extracolsep{0ptplus1fil}}}
		\hline\hline
		$\langle (2x-1)^m \rangle$    & & $m=2$ & $4$ & $6$ & $8$ & $10$ \\\hline
		This work & $\|$ & 0.20 & 0.087 & 0.048 & 0.031 & 0.022 \\
		& $\perp$ & 0.25 & 0.13 & 0.079 & 0.055 & 0.042 \\\hline
		$\varphi = \varphi^{\rm asy}$
		& & 0.20 & 0.086 & 0.048 & 0.030 & 0.021 \\
		$\varphi =\,$constant
		& & 0.33 & 0.2 & 0.14 & 0.11 & 0.091\\
		\hline
		\cite{PhysRevD.94.074018}
		& $\|$ & 0.25 & 0.12 & 0.075 & 0.052 & 0.038\\
		& $\perp$ & 0.26 & 0.13 & 0.079 & 0.054 & 0.039\\\hline
		\cite{Zhong:2023cyc}
		& $\|$ & 0.22 & 0.103 & 0.066 & 0.046 & 0.035\\
	%	& $\perp$ & 0.26 & 0.13 & 0.079 & 0.054 & 0.039
	    \hline
		\cite{Forshaw:2010py} %% AdS/QCD
		& $\|$ & 0.23 & 0.11 & 0.062 & 0.041 & 0.029\\
		& $\perp$ & 0.26 & 0.13 & 0.079 & 0.054 & 0.039\\\hline%
		\cite{Ball:1998sk} %% SR
		& $\|$ & 0.26 &  &   &   &   \\
		& $\perp$ & 0.27 &  &   &   &   \\\hline
		\cite{Bakulev:1998pf} %% nonlocal condensates
		& $\|$ & 0.23(1) & 0.095(5)& 0.051(4) &0.030(2)& 0.020(5)\\
		& $\perp$ & 0.33(1) &  &   &   &   \\\hline
		\cite{Pimikov:2013usa} %% nonlocal condensates
		& $\|$ & 0.22(2) & 0.089(9)& 0.048(5) &0.030(3)& 0.022(2)\\
		& $\perp$ & 0.11(1) & 0.022(2) &   &   &   \\\hline
		\cite{PhysRevD.75.034019} %% LC QM
		& $\|$ & 0.20(1) & 0.085(5) & 0.045(5) &  &  \\
		& $\perp$ & 0.21(1) & 0.095(5) & 0.05(1) &  &  \\\hline
		\cite{QCDSF-UKQCD:2007xbk,PhysRevD.83.074505}
		&  & 0.25(2)(2)  &  &   &   &   \\\hline
\hline	\end{tabular*}
\end{table}
%
%%%%%%%%%%%%%%%%%%%%%%%%%%%%%%%%
\subsection{$\rho$-meson form factors}
%%%%%%%%%%%%%%%%%%%%%%%%%%%%%%%%%
The LFWFs also provide direct access to electromagnetic form factors. 
The Lorentz-invariant electromagnetic Form factors $F_{i}\,(i=1,2,3)$ for a vector meson (spin-1) can be obtained by calculating the matrix elements of the electromagnetic current $J^{\mu}$ as~\cite{Arnold:1979cg,PhysRevD.70.053015},
\begin{align}\nonumber
    \langle V(P^\prime,\Lambda^\prime)\,|J^{\mu}|\,V(P,\Lambda)\rangle=&-\epsilon^{\ast}_{\Lambda^{\prime}}\cdot\epsilon_{\Lambda}(P+P^{\prime})^{\mu}F_{1}(Q^{2})  +\left(\epsilon^{\mu}_{\Lambda}q\cdot\epsilon^{\ast}_{\Lambda^{\prime}}-\epsilon^{\ast\mu}_{\Lambda^{\prime}}q\cdot\epsilon_{\Lambda^{\prime}}\right)F_{2}(Q^{2})\\
    &+\frac{(\epsilon^{\ast}_{\Lambda^{\prime}}\cdot q)(\epsilon_{\Lambda}\cdot q)}{2M_{V}^{2}}(P+P^{\prime})^{\mu}F_{3}(Q^{2})
\end{align}
where $\epsilon_{\Lambda}$ and $\epsilon_{\Lambda^{\prime}}$ are the polarization vectors of the initial and final mesons, respectively. We employ the Breit frame, where the momentum transfer occurs only in one transverse direction, i.e., ($q^{+}=0, \,  q_{x}=Q,\, q_{y}=0$), and $P_{\perp}=-P_{\perp}^{\prime}$~\cite{Cardarelli:1994yq,Brodsky:1992px}.  The momenta of the initial and final states are defined  as: $P^{\mu}=(M_{V}\sqrt{1+\eta},\,M_{V}\sqrt{1+\eta},\,-Q/2,0)$ and $P^{\prime\mu}=(M_{V}\sqrt{1+\eta},\,M_{V}\sqrt{1+\eta},\,Q/2,0)$, respectively 
with $\eta=Q^{2}/4M_{V}^{2}$. %with $P^{\mu}=(P^{+},P^{-},P^{\perp})$. 
We follow the notation $p^\mu=(p^+,\,p^-,\,p^1,\,p^2)$.
We compute the form factors by considering the plus component of the electromagnetic current,  $J^+(0)$. The matrix elements of  $J^+(0)$ can be expressed as~\cite{PhysRevC.102.055207,Li:2021cwv} 
\begin{align}
 I_{\Lambda^\prime, \Lambda}^{+} (Q^2) &\triangleq \langle V(P^\prime,\Lambda^\prime)| \frac{J^+(0)}{2P^+}| V(P,\Lambda) \rangle 
 \nonumber \\
 &= \sum_{h, \bar{h}} \int_0^1\int_0^{\infty} \frac{{\mathrm d}x{\mathrm d}^2\textbf{k}_\perp}{16\pi^{3}}
\Psi^{\Lambda^\prime*}_{h\bar{h}}(x,\mathbf{k}_\perp+(1-x)\textbf{q}_\perp)  \Psi^{\Lambda}_{h\bar{h}}(x,\mathbf{k}_\perp) ,\label{eq:FFs}
\end{align}
where $\Lambda$ and $\Lambda^\prime$ denote the helicities of the incoming and outgoing vector mesons, respectively. There are a total of nine matrix elements of the electromagnetic current, $I^{+}_{\Lambda\Lambda^{\prime}}$ for $\Lambda,\Lambda^{\prime}=0,\pm 1$. Using the light-front parity and time reversal invariance, one can reduce it to only four matrix elements: $I^{+}_{1,1}, I^{+}_{1,-1}, I^{+}_{1,0}$, and $I^{+}_{0,0}$. Note that the physical charge ($G_{C}$), magnetic ($G_{M}$), and quadrupole ($G_{Q}$) form factors are often employed to describe the electromagnetic properties of a hadron, instead of the Lorentz invariant electromagnetic form factors, $F_{i}$. However, these two types of form factors are related to each other such that
\begin{align}\nonumber
    G_{C}=F_{1}+\frac{2}{3}\eta G_{Q}\,, \quad\quad
    G_{M}=-F_{2}\,, \quad \quad
    G_{Q}=F_{1}+F_{2}+(1+\eta)F_{3}.
\end{align}
We obtain the static charge ($e$), magnetic moment ($\mu$) and quadrupole moment ($\mathcal{Q}$) of the hadron from the above form factors at zero momentum transfer, 
\begin{align}\nonumber
    eG_{\mathrm{C}}(0)=e\,,\quad\quad
    eG_{\mathrm{M}}(0)=2M_{V}\mu\,,\quad\quad
    -eG_{\mathrm{Q}}=&M_{V}^{2}\mathcal{Q}.
\end{align}
Notably, there are different prescriptions, for example, Grach and Kondratyuk (GK)~\cite{Grach:1983hd}, and Brodsky and Hiller (BH)~\cite{PhysRevD.46.2141}, to calculate such type of form factors. 
Nevertheless, we compute these physical form factors following the BH prescription, which includes the zero-mode contributions. In the BH prescription, the form factors are defined as,%~\cite{Grach:1983hd,Cardarelli:1994yq}
\begin{equation}
  \begin{aligned}\label{eq:ff_vector}
  G_{\mathrm{C}}^{\rm BH}(Q^2) &= \frac{1}{2P^{+}(1+2\eta)} \left[\frac{3-2\eta}{3}I_{0, 0}^{+} +\frac{16}{3}\eta \frac{I_{1,0}^{+}}{\sqrt{2\eta}}+\frac{2}{3}(2\eta-1)I_{1,-1}^{+}\right ],  \\
  G_{\mathrm{M}}^{\rm BH}(Q^2) &= \frac{2}{2P^{+}(1+2\eta)}\left[I_{0,0}^{+}+\frac{(2\eta-1)}{\sqrt{2\eta}}I_{1,0}^{+}-I_{1,-1}^{+}\right], \\
  G_{\mathrm{Q}}^{\rm BH}(Q^2) &= -\frac{1}{2P^{+}(1+2\eta)}\left[I_{0,0}^{+}-2\frac{I_{1,0}^{+}}{\sqrt{2\eta}}+\frac{1+\eta}{\eta}I_{1,-1}^{+} \right].  
  \end{aligned}
\end{equation}

\begin{figure}
		\centering	
	\includegraphics[scale=0.4]{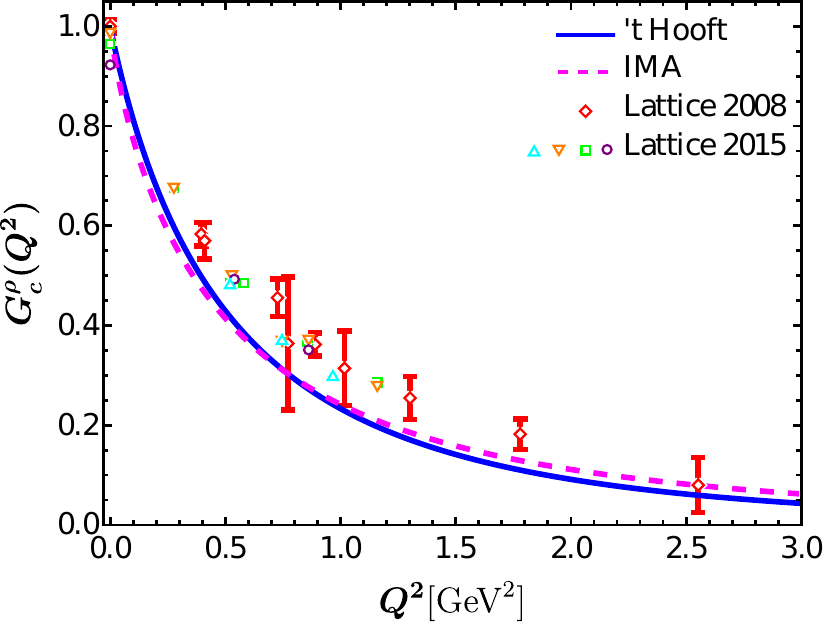}\hspace{0.1cm}
	\includegraphics[scale=0.4]{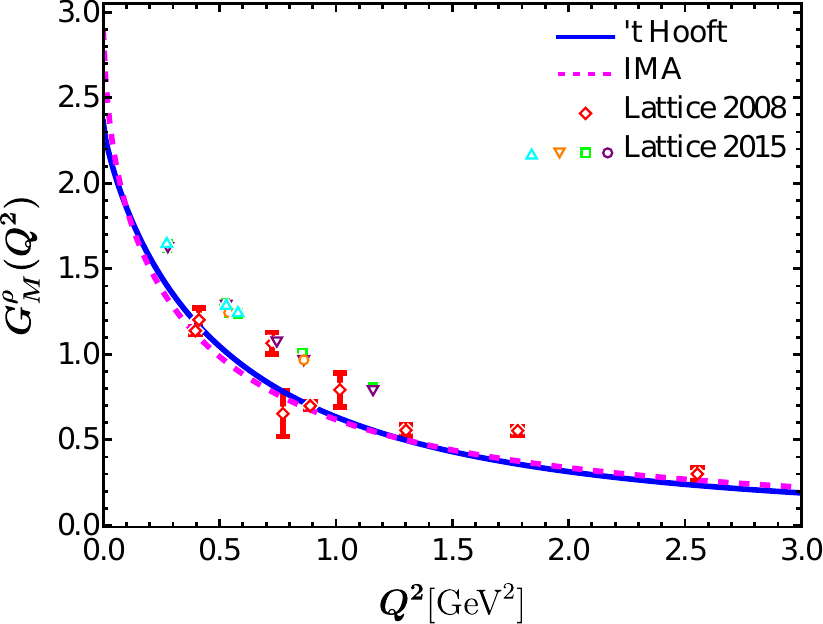}\hspace{0.1cm}
    \includegraphics[scale=0.4]{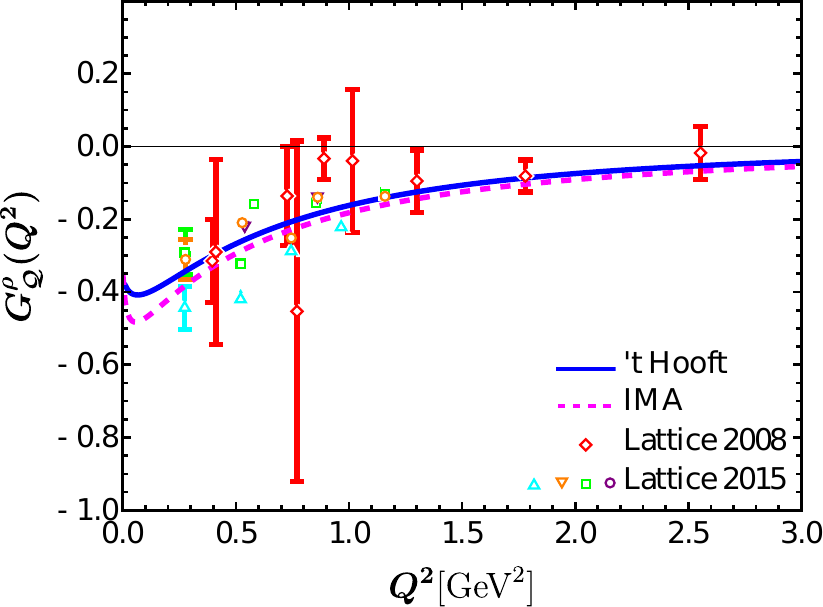}	
    \caption{Left: the charge $(G_{C}^{\rho})$; middle: the magnetic  $(G_{M}^{\rho})$; and right:  the quadrupole  $(G_{\mathcal{Q}}^{\rho})$ form factors for the $\rho$-meson as a functions of $Q^2$. Our results are compared with Lattice QCD predictions~\cite{QCDSF:2008tjq,Shultz:2015pfa}.}
	\label{Fig:EMFFs}
\end{figure}

 We show the variation of the the charge, magnetic, and quadrupole elastic form factors with $Q^{2}$ in Fig.~\ref{Fig:EMFFs}, where we  include the results generated using the IMA and Lattice QCD~\cite{QCDSF:2008tjq,Shultz:2015pfa} for comparison. %We observe that the qualitative behavior of the form factors in both approaches is consistent with each other. 
 We observe a good agreement of our results with the latice QCD simulations.
 From the charge form factor, we further calculate the charge root-mean-squared (rms) radius of the $\rho$ meson, which is defined as~\cite{Chung:1988my},
\begin{equation}
  \begin{aligned}\label{eq:ff_obs}
  \langle r^2_{\rho} \rangle &=  -\frac{6}{G_{\mathrm{C}}(0)}  \lim_{Q^2 \to 0}\frac{\partial G_{\mathrm{C}}(Q^2)}{\partial Q^2}.
  \end{aligned}
\end{equation}
We present our results for the static properties of the $\rho$ meson: rms charge radius, magnetic moment, and quadrupole moment 
in Table~\ref{tab:ff_obs1}, where we compare them 
with the predictions from various
theoretical approaches. We observe that our  result for the charge radius is close to the results from BSE~\cite{PhysRevC.77.025203}, lattice QCD~\cite{PhysRevD.91.074503}, and NJL model~\cite{PhysRevC.92.015212}. On the other hand, our magnetic moment is more or less consistent with all other studies summarized in Table~\ref{tab:ff_obs1}. The quadrupole moment %denoted as $\mathcal{Q}_{\rho}$, 
agrees well with the results from BSE~\cite{PhysRevC.77.025203}, LFQM~\cite{PhysRevD.70.053015}, and lattice QCD~\cite{PhysRevD.91.114501}, and they differ from other
predictions.
\begin{table*}
	%\centering
	\caption{Comparison of the $\rho$-meson charge radii $\sqrt{\langle r^2_{\rho} \rangle}$ (in $\text{fm}$), magnetic moments $\mu_{\rho}$ {(in units of Bohr magneton)} and quadrupole moments $\mathcal{Q}_{\rho}$ (in $\text{fm}^2$) with various theoretical approaches. } 
	\begin{ruledtabular}
		\begin{tabular}{c@{\hskip 0.1in}  c@{\hskip 0.1in}  c@{\hskip 0.1in}  c@{\hskip 0.1in}  c@{\hskip 0.1in}  c@{\hskip 0.1in}  c@{\hskip 0.1in} c@{\hskip 0.1in} c@{\hskip 0.1in}}
		 
			& This work 
            & LFH-IMA
			& BLFQ~\cite{PhysRevC.102.055207} 
			& BSE~\cite{PhysRevC.77.025203}    
			& Lattice QCD~\cite{PhysRevD.91.074503}
            & Lattice QCD~\cite{PhysRevD.91.114501}
			& LFQM~\cite{PhysRevD.70.053015}
			& NJL model~\cite{PhysRevC.92.015212}    
			\\%[3pt]
			\hline %\\[-6pt]
			$\sqrt{\langle r^2_{\rho}\rangle}$    & $0.75$
            & $0.94$    & $0.44$    &  $0.73$         & $0.819(42)$ & $0.55(5)$    & $0.52$      & $0.82$  \\%[4pt]
			$\mu_\rho$             & $2.40$
            & $2.90$    & 2.15  & 2.01         & 2.067(76) & 2.17(10)     & 1.92      & 2.48  \\ %[4pt]
			$\mathcal{Q}_\rho$  & $-0.027$
            &$-0.023$   & -0.063  & -0.026       & -0.0452(61) &-0.035  & -0.028    & -0.070 \\%[5pt]
		\end{tabular}
	\end{ruledtabular}
	\label{tab:ff_obs1}
\end{table*}

%%%%%%%%%%%%%%%%%
\section{Conclusion}\label{sec:conclusion}
%%%%%%%%%%%%%%%%%

  The 't Hooft equation is complementary to the light-front holographic Schr\"odinger equation, in governing the  longitudinal dynamics of quark-antiquark mesons. We have shown that together, they predict remarkably well the mass spectroscopy of $\rho$-meson family without further adjusting parameters: the universal transverse confinement scale  $\kappa=0.523$ GeV, the longitudinal confinement scale $g=0.109$ GeV, and the light quark mass $m_q=0.046$ GeV, which were determined to predict the pion spectroscopy and its structure~\cite{Ahmady:2022dfv}. In conjunction with the color glass condensate dipole cross-section, the $\rho$-meson holographic LFWFs  after incorporating the longitudinal mode generated by the ’t Hooft equation lead to a good description of the cross-section data
for the diffractive $\rho$-meson electroproduction at different energies. Using the resulting LFWFs, we have calculated the decay constant,  distribution amplitude,
 electromagnetic form factors,  charge radius, magnetic moment, and quadrupole moment of the $\rho$-meson. Interestingly, we have noticed that although, the electromagnetic form factors in our approach agree well with the LFH-IMA predictions, they differ from each other in describing the distribution amplitudes. We have found that the vector coupling is close to the experimentally measured data and various theoretical predictions; however, the tensor coupling constant is significantly smaller compared
to the other predictions in the literature. Meanwhile, the moments of distribution amplitudes and the static properties:  charge radius, magnetic moment and quadrupole moment have been found to be consistent with other theoretical as well as lattice QCD results.%, but charge radius strongly deviates from other theoretical predictions.

%%%%%%%%%%%%%%%%%
\section*{Acknowledgement}
%%%%%%%%%%%%%%%%%
We thank Ruben Sandapen and Mohammad Ahmady for fruitful discussions. C.M. is supported by new faculty start up funding by the Institute of Modern Physics, Chinese Academy of Sciences, Grant No.
E129952YR0. C.M. also thanks the Chinese Academy of
Sciences Presidents International Fellowship Initiative for the support via Grant No. 2021PM0023. S.K. is supported by Research Fund for
International Young Scientists, Grant No. 12250410251, from the National Natural Science Foundation of China
(NSFC), and China Postdoctoral Science Foundation
(CPSF), Grant No. E339951SR0.

\bibliography{ref.bib}

\end{document}